\documentclass[aps,prl,floatfix,superscriptaddress,preprintnumbers,nofootinbib,11pt]{revtex4-1}
\RequirePackage[colorlinks=true, urlcolor=blue, anchorcolor=blue, citecolor=blue, filecolor=blue, linkcolor=blue, menucolor=blue, linktocpage=true, pdfproducer=medialab, pdfa=true]{hyperref}
\usepackage[utf8]{inputenc}
\usepackage{epsfig,latexsym,cancel,amssymb,amsmath,verbatim,mathrsfs}
\usepackage{xcolor}
\usepackage{subfigure}
\usepackage{graphicx}
\usepackage{bm}
\usepackage{multirow}
\usepackage[title,titletoc]{appendix}

\definecolor{ccblue}{rgb}{0.0,0.4,0.8}
\usepackage[]{hyperref}
\hypersetup{
  colorlinks=true,
  linkcolor=ccblue,
  urlcolor=ccblue,
  citecolor=ccblue
}

\begin{document}

\title{Confronting cosmic ray electron and positron excesses with hybrid triplet Higgs portal dark matter}
\author{Shao-Long Chen}
\email[E-mail: ]{chensl@mail.ccnu.edu.cn}
\affiliation{Key Laboratory of Quark and Lepton Physics (MoE) and Institute of Particle Physics, Central China Normal University, Wuhan 430079, China}
\affiliation{Center for High Energy Physics, Peking University, Beijing 100871, China}

\author{Amit Dutta Banik}
\email[E-mail: ]{amitdbanik@mail.ccnu.edu.cn}
\affiliation{Key Laboratory of Quark and Lepton Physics (MoE) and Institute of Particle Physics, Central China Normal University, Wuhan 430079, China}

\author{Ze-Kun Liu}
\email[E-mail: ]{zekunliu@mails.ccnu.edu.cn}
\affiliation{Key Laboratory of Quark and Lepton Physics (MoE) and Institute of Particle Physics, Central China Normal University, Wuhan 430079, China}


\begin{abstract}
We perform a detailed study of scalar dark matter with triplet Higgs extensions of the Standard Model in order to explain the cosmic ray electron and positron excesses reported by AMS-02 and DAMPE. A detailed analysis of AMS-02 positron excess reveals that for different orderings (normal, inverted and quasi-degenerate) of neutrino mass, the hybrid triplet Higgs portal framework is more favored with respect to the single triplet Higgs portal for TeV scale dark matter. We also show that the resonant peak and continuous excess in DAMPE cosmic ray data can be well explained with the hybrid triplet Higgs portal dark matter when a dark matter sub-halo nearby is taken into account. 
\end{abstract}

\maketitle

\section{Introduction} 
The existence of dark matter (DM) is well-established by cosmological and astrophysical evidences~\cite{ParticleDataGroup:2020ssz}. However, the nature of DM still remains mysterious. Cosmic ray electrons and positrons (CRE) originating from adjacent galactic sources are smoking gun probes to signature of DM annihilation or decay. Different terrestrial and satellite based experiments, such as HESS \cite{Aharonian:2008aa,Aharonian:2009ah}, Fermi-LAT~\cite{Abdollahi:2017nat,Meehan:2017xlr}, VERITAS \cite{Holder:2016heb,Staszak:2015lga}, AMS-02~\cite{AMS:2013fma, AMS:2014bun, AMS:2018avs} etc., measure cosmic ray electron and positron flux up to TeV scale energy. The AMS-02 experiment show distinctive features of positron flux data with a low energy excess around $E\sim 25$ GeV and excess with a peak at about 300 GeV compared to power-law background, which can be well described by the background with an extra positron source~\cite{Aguilar:2019owu}.  DAMPE satellite borne experiment~\cite{Ambrosi:2017wek} also reported their measurements of cosmic ray electron/position spectrum ranging from 25 GeV to 4.6 TeV. DAMPE CRE data~\cite{TheDAMPE:2017dtc} reports a spectral break around 0.9 TeV, which can't be fitted properly by the single power-law background, and an excess peak around 1.5 TeV which is expected to have possible origin from DM annihilation or (and) decay. The most convincing explanation to this 1.5 TeV excess is based on the assumption that the excess peak is due to DM annihilation (and/or decays) in nearby sub-halo enriched with DM.  Various studies have been extensively performed to explain the mentioned electron/position excesses and the possible flavor composition of DM annihilation final states~\cite{Liu:2017rgs, Yuan:2017ysv, Ding:2017jdr, Huang:2017egk, Feng:2019rgm, Coogan:2019uij, Liu:2019iik, Ge:2017tkd, Ghosh:2020fdc, Ge:2020tdh, Ding:2020wyk}.

In the present work, we consider a scalar DM accompanied by type-II seesaw motivated triplet Higgs extension of the Standard Model (SM), in order to explain AMS-02 and DAMPE cosmic ray electron and positron spectrum. Triplet scalars dominantly decay into leptons with their vacuum expectation values (VEVs) less than $10^{-5}$ GeV, therefore they could play a perfect role as a leptonic portal~\cite{Chen:2009ew, Gogoladze:2009gi, Dev:2013hka} and provide a viable neutrino mass origin mechanism simultaneously. The scalar DM candidate is protected by unbroken $Z_2$ symmetry, couples to Higgs triplet via quartic interaction which provides the necessary leptonic annihilations. Earlier attempts to explain CRE obtained from AMS-02 and DAMPE data are carried out with single triplet Higgs extension of the SM with scalar DM~\cite{Dev:2013hka, Li:2018abw, Li:2017tmd, Sui:2017qra}. In the present work, we consider a hybrid triplet Higgs extension of the SM. We show that such hybrid triplet Higgs portal DM model can provide better fit to recent AMS-02 positron excess compared with the single triplet Higgs extension of the SM. We consider the DAMPE data with a single power-law background and there exists a significant excess cosmic ray electron/positron spectrum ranging from 20 GeV to 1.5 TeV along with the excess peak-like feature. We show that in the hybrid triplet Higgs portal model, the reported excesses in AMS-02/DAMPE data are well fitted when electron/positron excess from Milky Way (MW) and nearby DM sub-halo is taken into account. We find that there is common parameters space to fit both AMS-02 and DAMPE data. We perform our analysis considering all possible neutrino mass hierarchies (normal, inverted and quasi-degenerate) and the corresponding flavor structure is explored in detail.

The paper is organized as follows: We first briefly discuss both single and hybrid triplet Higgs portal model and neutrino Yukawa couplings are derived for different mass ordering of neutrinos (normal, inverted and quasi-degenerate) by neutrino oscillation experimental data. In the next section, we describe the production and propagation of cosmic ray electron and positron flux within the galaxy. We then perform detailed analysis of AMS-02 positron excess data with single and hybrid triplet Higgs models separately for different neutrino mass hierarchies and acquire the parameters space for fitting the AMS-02 data. We show that hybrid triplet Higgs portal model can provide better explanation for AMS-02 results. In the next section, we provide the fit results of DAMPE excess arising from MW and DM sub-halo nearby, and common parameter space to fit both AMS-02 and DAMPE data. Finally we summarize the work with concluding remarks.

\section{The model}
\subsection{The single triplet Higgs portal scenario}
In addition to the SM fields, an extra $\rm{SU(2)_L}$ triplet Higgs field is introduced, providing a solution to the neutrino mass origin which is the so called type-II seesaw mechanism. We denote triplet complex scalar fields $\Delta$ with hypercharge $Y=2$ and SM doublet Higgs $\Phi$ as
\begin{equation}
\Delta(3,+2)=\begin{pmatrix}
\frac{1}{\sqrt{2}}\Delta^+ & \Delta^{++}\\
\Delta^0 & -\frac{1}{\sqrt{2}}\Delta^+
\end{pmatrix}\,,
\qquad
\Phi(2,+1)=\begin{pmatrix}
\phi^+ \\ \phi^0
\end{pmatrix}\,,
\end{equation}
where $\Delta^0 = (\delta + i \eta)/\sqrt{2} $ and $\phi^0 = (\phi_R + i \phi_I)/\sqrt{2}$. The most general scalar potential involving $\Delta$ and $\Phi$ can be written as
\begin{equation}
\begin{split}
V(\Delta ,\,\Phi)=&m_{\phi}^2 \,\Phi^{\dagger}\Phi + m_{\Delta}^2 \text{Tr} (\Delta^{\dagger} \Delta) + \rho_{\phi} \,(\Phi^{\dagger}\Phi)^2 + \frac{1}{2} \rho_1 \,[\text{Tr} (\Delta_k^{\dagger} \Delta_l)]^2  +  \frac{1}{2}\rho_2 \,\text{Tr} (\Delta^{\dagger} \Delta^{\dagger})\,\text{Tr} (\Delta \Delta)\\
&+ \sigma_1 \,\Phi^{\dagger}\Phi\,\text{Tr}(\Delta^{\dagger} \Delta) + \sigma_2 \,\Phi^{\dagger}\Delta^{\dagger} \Delta \Phi + (\mu\,\Phi^{\dagger}\Delta \tilde{\Phi}+ \text{h.c.})\,,
\end{split}
\label{1THMV}
\end{equation}
where $\tilde{\Phi}=i\tau_2 \Phi^*$. The VEVs of the Higgs fields are given by
\begin{equation}
\langle \Delta \rangle=\frac{1}{\sqrt{2}}
\begin{pmatrix}
0 & 0\\
v_{\Delta} & 0
\end{pmatrix}\,,
\qquad
\langle \Phi\rangle =\frac{1}{\sqrt{2}}
\begin{pmatrix}
0 \\  v
\end{pmatrix}\,.
\end{equation}
Without loss of generality, we assumed all parameters and VEVs are real. The triplet Higgs VEV is naturally chosen as $v_{\Delta} \sim \text{keV}$, well below the constraints from the electroweak precision test~\cite{delAguila:2008ks}. The conditions for minimization of potential are
\begin{align}
m_{\phi}^2 + \rho_{\phi} v^2 + (\frac{1}{2} \sigma_1 v_{\Delta} + \sqrt{2} \mu)\,v_{\Delta} &= 0\,,\\
m_{\Delta}^2 v_{\Delta} + \rho_1 v_{\Delta}^3 + (\frac{1}{2} \sigma_1 v_{\Delta} + \sqrt{2} \mu)\, v^2 &= 0\,.
\end{align}
After Higgs fields $\Delta$ and $\Phi$ acquiring VEVs due to the spontaneous symmetry breaking, the mass squared matrices of CP-even Higgs fields $(\phi_R ,\delta)$, CP-odd Higgs fields $(\phi_I ,\eta)$,  singly-charged Higgs fields $(\phi^+ ,\Delta^+)$, double-charged Higgs fields $\Delta^{\pm\pm}$ are given respectively by
\begin{align}
(M^0_{R})^2&=
\begin{pmatrix}
m_{\phi}^2 + 3 \rho_{\phi} v^2 + \sqrt{2}\mu v_{\Delta} + \frac{1}{2}\sigma_1 v_{\Delta}^2 & \sigma_1 v v_{\Delta}+ \sqrt{2}\mu v \\
\sigma_1 v v_{\Delta}+ \sqrt{2} \mu v  &  m_{\Delta}^2 + \frac{1}{2} \sigma_1 v^2 + \frac{3}{2} \rho_1 v_{\Delta}^2\\
\end{pmatrix}\,,
\\
(M^0_{I})^2&=
\begin{pmatrix}
m_{\phi}^2 +  \rho_{\phi} v^2 - \sqrt{2}\mu v_{\Delta} + \frac{1}{2}\sigma_1 v_{\Delta}^2 & \sqrt{2} \mu v \\
\sqrt{2} \mu v & m_{\Delta}^2 + \frac{1}{2} \sigma_1 v^2 + \frac{1}{2}\rho_1 v_{\Delta}^2\\
\end{pmatrix}\,,\\
(M^+)^2&=
\begin{pmatrix}
m_{\phi}^2 + \rho_{\phi} v^2 + \frac{1}{2}(\sigma_1 + \sigma_2)v_{\Delta}^2& \mu v - \frac{\sqrt{2}}{4} \sigma_2 v v_{\Delta} \\
\mu v - \frac{\sqrt{2}}{4} \sigma_2 v v_{\Delta} & m_{\Delta}^2 + \frac{1}{2} \sigma_1 v^2+ \frac{1}{4}\sigma_2 v^2+ \frac{1}{2} \rho_1 v_{\Delta}^2\\
\end{pmatrix}\,, \\
(M^{++})^2&=m_{\Delta}^2 + \frac{1}{2}\sigma_1 v^2+ \frac{1}{2}\sigma_2 v^2+ \frac{1}{2} (\rho_1 + 2 \rho_2)v_{\Delta}^2\,.
\end{align}
By diagonalizing the above mass squared matrices, we get seven physical states: CP-even Higgs boson $h$ and $H$, CP-odd Higgs boson $A$, singly-charged Higgs boson $H^{\pm}$ and double-charged Higgs boson $H^{\pm\pm}$. The coupling parameters in the scalar potential are expressed in terms of the masses of physical particles and mixing angle $\alpha$ between two CP-even Higgses,
\begin{align}
&\rho_{\phi} = \frac{1}{v^2}\left(m_{h}^2c_\alpha^2 + m_{H}^2 s_\alpha^2 \right) \,,\\
&\rho_1 = \frac{1}{2 v_{\Delta}^3}\left(\sqrt{2} \mu v^2 + 4 m_{h}^2 s_\alpha^2 v_{\Delta} + 4 m_{H}^2 c_\alpha^2 v_{\Delta}\right)\,,\\
&\rho_2 = \frac{1}{2 v_{\Delta}^3}\left(\sqrt{2} \mu v^2 + 2m_{H^{\pm\pm}}^2 v_{\Delta} -\sigma_2 v^2 v_{\Delta} \right) \,,\\
&\sigma_1 = \frac{1}{vv_{\Delta}}\left(2\left(m_{h}^2 - m_{H}^2 \right)  s_\alpha c_\alpha- \sqrt{2} \mu v \right) \,,\\
&\sigma_2 = \frac{1}{v^4 + 6 v^2 v_{\Delta}^2 + 8 v_{\Delta}^4}\left(4m_{H^{\pm}}^2( v^2 + 4 v_{\Delta}^2) - 8 m_A^2(v^2 + 2 v_{\Delta}^2) \right) \,,\\
&\mu = -\frac{2\sqrt{2}m_A^2 v_{\Delta}}{v^2+4v_{\Delta}^2}\,,
\end{align}
where we denote $c_\alpha\equiv\cos \alpha$ and $s_\alpha\equiv\sin \alpha$. The triplet field $\Delta$ can couple with a pair of lepton doublet  $L_L^i=(\nu_L^i,\ell_L^i)^T$  through the Yukawa terms, which are written as
\begin{align} \nonumber
\mathcal{L}_{\Delta}=&\frac{1}{\sqrt{2}}(Y_{\Delta})_{ij}\overline{(L_L^i)^C}i\tau_2\Delta L_L^j + \text{h.c.}\\ \nonumber
=&\frac{1}{\sqrt{2}}(Y_{\Delta})_{ij}\overline{(\nu_L^i)^C}\Delta^0 \nu_L^j - \frac{1}{2}(Y_{\Delta})_{ij}\overline{(\nu_L^i)^C}\Delta^+ \ell_L^j - \frac{1}{2}(Y_{\Delta})_{ij}\overline{(\ell_L^i)^C}\Delta^+ \nu_L^j\\
&- \frac{1}{\sqrt{2}}(Y_{\Delta})_{ij}\overline{(\ell_L^i)^C} \Delta^{++} \ell_L^j + \text{h.c.},
\label{deltaY}
\end{align}
where $C$ denotes the charge conjugation. When the triplet Higgs field $\Delta$ acquires VEV, the lepton number is broken and the neutrino mass matrix then is generated from Eq.~(\ref{deltaY})
\begin{equation}
(M_{\nu})_{ij}=(Y_{\Delta})_{ij}v_{\Delta}\,.
\end{equation}
The Yukawa coupling matrix $Y_{\Delta}$ is fixed by the three neutrino masses and the Pontecorvo-Maki-Nakagawa-Sakata (PMNS) mixing matrix $U_{\text{PMNS}}$
\begin{equation}
Y_{\Delta}=\frac{1}{v_{\Delta}}U^*_{\text{PMNS}} M_{\nu}^{\text{diag}} U^{\dagger}_{\text{PMNS}}\,,
\label{Yukawa1}
\end{equation}
where $M_{\nu}^{\text{diag}}$ denotes the diagonal neutrino mass matrix. The PMNS matrix can be parameterized as~\cite{ParticleDataGroup:2020ssz}
\begin{equation}
U_{\text{PMNS}}=\begin{pmatrix}
c_{12} c_{13} & s_{12} c_{13}  & s_{13}e^{-i\delta} \\
-s_{12}c_{23}-c_{12}s_{23}s_{13}e^{i\delta}  & c_{12}c_{23}-s_{12}s_{23}s_{13}e^{i\delta} & s_{23}c_{13} \\
s_{12}s_{23}-c_{12}c_{23}s_{13}e^{i\delta} & -c_{12}s_{23}-s_{12}c_{23}s_{13}e^{i\delta} & c_{23} c_{13} 
\end{pmatrix}\,,
\end{equation}
where $c_{ij}\equiv \cos\theta_{ij}$, $s_{ij}\equiv\sin\theta_{ij}$, $\delta$ is the the Dirac CP phase. For simplicity, we have
set the Majorana phases in the PMNS matrix to be zero. 
\begin{table}[ht]
 \begin{center}
 \begin{tabular}{|c|c|c|c|}
 \hline
 Quantity    & Normal ordering & Inverted ordering\\ \hline
 $\theta_{12} /^{\circ}$ & 33.44 & 33.45 \\ \hline
 $\theta_{23} /^{\circ}$ & 49.0 & 49.3 \\ \hline
 $\theta_{13} /^{\circ}$ & 8.57 & 8.61 \\ \hline
 $\delta_{\text{CP}} /^{\circ}$ & 195 & 286 \\ \hline
 $\Delta m_{21}^2\ [\text{eV}^2]$ & $7.42 \times 10^{-5}$ & $7.42 \times 10^{-5}$ \\ \hline
 $\Delta m_{31}^2\ [\text{eV}^2]$ & $2.514 \times 10^{-3}$ & $-2.497 \times 10^{-3}$ \\ \hline
 \end{tabular}
 \end{center}
 \caption{Global fit of neutrino oscillation parameters, taken from~\cite{Esteban:2020cvm}.}
 \label{nupara}
\end{table}
The values of neutrino oscillation data we used are listed in Table~\ref{nupara}. The cosmological upper limit on the sum of three light neutrino masses is given by the Planck data at $95\%$ confidence level, $\sum m_{i} \leq 0.12\ \text{eV}$~\cite{Planck:2018vyg}. The exact values of each neutrino mass and even the mass ordering still remain unknown. In this work, we have set the lightest neutrino mass to be zero for both normal hierarchy (NH) and inverted hierarchy (IH), and then the other two neutrino masses can be fixed by the oscillation data. For the quasi-degenerate hierarchy (DH) case, we assume that the sum of the three neutrinos masses is 0.12 eV. After confirming the neutrinos masses, and the Yukawa coupling matrix in the Eq.~(\ref{Yukawa1}) for three different hierarchy cases are given respectively by
\begin{align}
Y_{\Delta}&=\dfrac{10^{-2}\, \text{eV}}{v_{\Delta}}\begin{pmatrix}
0.35+0.06\,i & -0.25-0.14\,i  & -0.74-0.12\,i \\
-0.25-0.14\,i & 3.11+0.02\,i & 2.13-0 \,i \\
 -0.74-0.12\,i & 2.13-0 \,i & 2.40-0.01\,i 
\end{pmatrix}
\quad \text{(NH)}\,,
\label{YNH}
\\
Y_{\Delta}&=\dfrac{10^{-2}\, \text{eV}}{v_{\Delta}}\begin{pmatrix}
4.92+0\,i & 0.56-0.14\,i  & 0.45-0.13\,i \\
0.56-0.14\,i & 2.24+0.03\,i & -2.46-0.03 \,i \\
0.45-0.13\,i & -2.46-0.03 \,i & 2.92-0.02\,i 
\end{pmatrix}
\quad \text{(IH)}\,,
\label{YIH}
\\
Y_{\Delta}&=\dfrac{10^{-2}\, \text{eV}}{v_{\Delta}}\begin{pmatrix}
7.14+0.10\,i & -0.15-0.46\,i & -0.16-0.40\,i \\
-0.15-0.46\,i & 8.01+0.05\,i & 0.74+0.04 \,i \\
-0.16-0.40\,i & 0.74+0.04\,i & 7.80+0.03\,i
\end{pmatrix}
\quad \text{(DH)}\,.
\label{YDH}
\end{align}

The kinetic terms in the Lagrangian for the SM doublet $\Phi$ and triplet $\Delta$ are given by
\begin{equation}
\mathcal{L}_{\text{gauge}}=(D_{\mu}\Phi)^{\dagger}D^{\mu}\Phi + \text{Tr}\,[(D_{\mu}\Delta)^{\dagger}(D^{\mu}\Delta)]\,,
\end{equation}
where $D_{\mu}\Phi=\partial_{\mu}\Phi-\frac{i}{2}gW_{\mu}^a\tau^a \Phi - \frac{i}{2}g^{\prime}B_{\mu}\Phi$ and $D_{\mu}\Delta=\partial_{\mu}\Delta-\frac{i}{2}g[W_{\mu}^a\tau^a, \Delta] - g^{\prime}B_{\mu}\Delta$. The doubly charged scalar $(H^{\pm\pm})$ can decay into a singly charged scalar and a $W$ boson $(H^{\pm}W^{\pm})$, a pair of lepton $(\ell^{\pm}\ell^{\pm})$ or a pair of $W$ boson $(W^{\pm}W^{\pm})$. The channel $H^{\pm\pm} \to H^{\pm} W^{\pm}$ can be kinematically forbidden by taking $m_{H^{\pm\pm}} \simeq m_{H^{\pm}}$. The branching ratio of channel $H^{\pm\pm} \to \ell^{\pm}\ell^{\pm}$ and channel $H^{\pm\pm} \to W^{\pm}W^{\pm}$ depends on the value of triplet Higgs VEV and the $H^{\pm\pm}$ dominantly decays into leptons when the triplet Higgs VEV $v_{\Delta} < 10^{-5}$ GeV~\cite{Dey:2020tfq}.
With the assumption of $m_{H^{\pm}} \simeq m_{H} \simeq m_{A}$ and $v_{\Delta} < 10^{-5}$ GeV,  $H^{\pm}$, $H$ and $A$ dominantly decay into $\ell^{\pm}\nu$, $\nu\nu$ and $\nu\nu$ respectively.

\subsection{The hybrid triplet Higgs portal scenario}
In this work, we consider a hybrid triplet portal model with two Higgs triplets to explain cosmic ray electron and positron excesses and compare this scenario with the single triplet Higgs model. Models with two triplet Higgs fields have been proposed to address neutrino mass matrix pattern and leptogenesis~\cite{Ma:1998dx, Chaudhuri:2013xoa, Parida:2020sng}, and was explored for the context of collider experiments~\cite{Chaudhuri:2013xoa,Ghosh:2018jpa}. We denote two complex triplet Higgs fields $\Delta_1$ and $\Delta_2$ with hypercharge $Y=2$ as 
\begin{equation}
\Delta_{1}(3,+2)=\begin{pmatrix}
\frac{1}{\sqrt{2}}\Delta_{1}^+ & \Delta_{1}^{++}\\
\Delta^0_{1} & -\frac{1}{\sqrt{2}}\Delta_{1}^+
\end{pmatrix}\,,
\qquad
\Delta_{2}(3,+2)=\begin{pmatrix}
\frac{1}{\sqrt{2}}\Delta_{2}^+ & \Delta_{2}^{++}\\
\Delta^0_{2} & -\frac{1}{\sqrt{2}}\Delta_{2}^+
\end{pmatrix}\,,
\end{equation}
where $\Delta_{1,2}^0 = (\delta_{1,2}^0 + i \eta_{1,2})/\sqrt{2}$. The general scalar potential of these two triplet Higgs fields $\Delta_1$, $\Delta_2$ and SM doublet Higgs fields $\Phi$ can be written as
\begin{align}\nonumber
V(\Delta_k ,\,\Phi)=&m_{\phi}^2 \,\Phi^{\dagger}\Phi + m_{kl}^2 \text{Tr} (\Delta_k^{\dagger} \Delta_l) + \rho_{\phi} \,(\Phi^{\dagger}\Phi)^2 + \frac{1}{2} \rho_{kl} \,[\text{Tr} (\Delta_k^{\dagger} \Delta_l)]^2  +  \frac{1}{2}\rho^{\prime}_{kl} \,\text{Tr} (\Delta_k^{\dagger} \Delta_l^{\dagger})\,\text{Tr} (\Delta_k \Delta_l)\\ \nonumber
& + \rho_{\Delta} \,\text{Tr} (\Delta_1^{\dagger} \Delta_1)\,\text{Tr} (\Delta_2^{\dagger} \Delta_2) + \rho_{\Delta}^{\prime} \,\text{Tr} (\Delta_1^{\dagger} \Delta_2)\,\text{Tr} (\Delta_2^{\dagger} \Delta_1) + \sigma_{kl}\,\Phi^{\dagger}\Phi\,\text{Tr}(\Delta_k^{\dagger} \Delta_l) \\ 
& + \sigma^{\prime}_{kl}\,\Phi^{\dagger}\Delta_k^{\dagger} \Delta_l \Phi+(\mu_k\,\Phi^{\dagger}\Delta_k \tilde{\Phi}+\text{h.c.})\,,
\end{align}
where $k,l=1,2$. The VEVs of the Higgs fields are given by
\begin{equation}
\langle \Delta_{1,2} \rangle=\frac{1}{\sqrt{2}}\begin{pmatrix}
0 & 0\\
v_{1,2} & 0
\end{pmatrix}\,.
\end{equation}
The orders of magnitude for the parameters in the potential are assumed:
\begin{equation}
m_{\phi}, m_{kl} \sim v; \qquad \rho_{\phi}, \rho_{kl}, \rho^{\prime}_{kl}, \rho_{\Delta}, \rho_{\Delta}^{\prime}, \sigma_{kl}, \sigma^{\prime}_{kl} \sim 1; \qquad |\mu_{k}| \ll 1; \qquad v_1,v_2 \ll v\,.
\end{equation}
For convenience, we define some matrices as
\begin{equation}
\begin{split}
&\bm{m}=(m_{kl})\,, \qquad \bm{\rho}=(\rho_{kl})\,, \qquad \bm{\rho^{\prime}}=(\rho^{\prime}_{kl})\,, \qquad \bm{\sigma}=(\sigma_{kl})\,, \qquad \bm{\sigma^{\prime}}=(\sigma^{\prime}_{kl})\,, \\[0.3cm]
&\bm{\mu}=\dbinom{\mu_1}{\mu_2}\,, \qquad \bm{V}=\dbinom{v_1}{v_2}\,.
\end{split}
\end{equation}
With the approximation of $v_1,v_2 \ll v$, the conditions for minimization of potential are
\begin{equation}
\begin{split}
m_{\phi}^2 + \rho_{\phi}v^2 + \sqrt{2}\bm{\mu}^T \bm{V}=&0\,,\\
(\bm{m}^2+\frac{v^2}{2}\bm{\sigma})\bm{V}+\frac{\sqrt{2}}{2}\bm{\mu}v^2=&0\,.
\end{split}
\end{equation}
The mass squared matrices of CP-even Higgs $(\phi_R ,\delta_1, \delta_2)$, CP-odd Higgs $(\phi_I ,\eta_1, \eta_2)$,  singly-charged Higgs $(\phi^+ ,\Delta_1^+, \Delta_2^+)$, double-charged Higgs $(\Delta_1^{++}, \Delta_2^{++})$ are given respectively by
\begin{align}
(M^0_{R})^2_{3 \times 3}&=
\begin{pmatrix}
m_{\phi}^2 + 3 \rho_{\phi} v^2 + \sqrt{2}\mu_1v_1 + \sqrt{2}\mu_2 v_2 & (\bm{\sigma} \bm{V} v+ \sqrt{2}\bm{\mu} v )^T \\
\bm{\sigma} \bm{V} v+ \sqrt{2}\bm{\mu} v  & \bm{m}^2 + \frac{1}{2} \bm{\sigma} v^2\\
\end{pmatrix}\,,
\\
(M^0_{I})^2_{3 \times 3}&=
\begin{pmatrix}
m_{\phi}^2 + \rho_{\phi} v^2 - \sqrt{2}\mu_1v_1 - \sqrt{2}\mu_2v_2 & (\sqrt{2}\bm{\mu}v )^T \\
\sqrt{2}\bm{\mu}v & \bm{m}^2 + \frac{1}{2} \bm{\sigma} v^2\\
\end{pmatrix}\,,
\\
(M^+)^2_{3 \times 3}&=
\begin{pmatrix}
m_{\phi}^2 + \rho_{\phi} v^2& (\bm{\mu} v - \frac{\sqrt{2}}{4}v \bm{\sigma^{\prime}}\bm{V})^T \\
 \bm{\mu} v - \frac{\sqrt{2}}{4}v \bm{\sigma^{\prime}}\bm{V} & \bm{m}^2 + \frac{1}{2} \bm{\sigma} v^2+ \frac{1}{4} \bm{\sigma^{\prime}}v^2\\
\end{pmatrix}\,, \\
(M^{++})^2_{2 \times 2}&=\bm{m}^2 + \frac{1}{2} \bm{\sigma} v^2+ \frac{1}{2} \bm{\sigma^{\prime}}v^2\,.
\end{align}
After diagonalization of these matrices, we obtain three Goldstone bosons and thirteen physical states. The transformations are parametrized by
\begin{equation}
\begin{pmatrix}
h \\ H_1 \\ H_2
\end{pmatrix}=U_R
\begin{pmatrix}
\phi_R \\ \delta_1 \\ \delta_2
\end{pmatrix}\,,
\begin{pmatrix}
G^0 \\ A_1 \\ A_2
\end{pmatrix}=U_I
\begin{pmatrix}
\phi_I \\ \eta_1 \\ \eta_2
\end{pmatrix}\,,
\begin{pmatrix}
G^{\pm} \\ H^{\pm}_1 \\ H^{\pm}_2
\end{pmatrix}=U_P
\begin{pmatrix}  
\phi^{\pm} \\ \Delta_1^{\pm} \\ \Delta_2^{\pm}
\end{pmatrix}\,,
\begin{pmatrix}
H^{\pm\pm}_1 \\ H^{\pm\pm}_2 
\end{pmatrix}=U_{PP}
\begin{pmatrix}
\Delta^{\pm\pm}_1 \\ \Delta^{\pm\pm}_2
\end{pmatrix}\,.
\end{equation}
The Yukawa terms for triplet Higgses $\Delta_1, \Delta_2$ coupled to leptons are given by
\begin{equation}
\mathcal{L}_{\Delta}^{\prime}=\frac{1}{\sqrt{2}} (Y_{\Delta_1})_{ij} \overline{(L_L^i)^C}i\tau_2\Delta_1 L_L^j + \frac{1}{\sqrt{2}} (Y_{\Delta_2})_{ij} \overline{(L_L^i)^C}i\tau_2\Delta_2 L_L^j + h.c..
\label{Yukawa2}
\end{equation}
When the triplet Higgs fields acquire VEVs, the neutrino mass matrix is generated,
\begin{equation}
(M_{\nu})_{ij} = (Y_{\Delta_1})_{ij} v_1 + (Y_{\Delta_2})_{ij} v_2 \,.
\end{equation}
Compared with the single triplet Higgs portal case, the coupling matrices $Y_{\Delta_1}$ and $Y_{\Delta_2}$ are more flexible after fixing the VEVs of triplet Higgses. Gauge interactions of the SM Higgs and triplet Higgses are governed by
\begin{equation}
\mathcal{L}_{\text{gauge}}=(D_{\mu}\Phi)^{\dagger}D^{\mu}\Phi + \text{Tr}\,\left[(D_{\mu}\Delta_1)^{\dagger}(D^{\mu}\Delta_1)\right]+ \text{Tr}\,\left[(D_{\mu}\Delta_2)^{\dagger}(D^{\mu}\Delta_2)\right]\,.
\end{equation}
The charged Higgs $H_{1,2}^{\pm}$ and $H_{1,2}^{\pm\pm}$ dominantly decay into leptons in the scenario with the parameters set as: small VEVs of triplet fields $v_{1,2} < 10^{-5}$ GeV, small mass splitting between two triplet Higgses $m_{H_{i}^{\pm\pm}} \simeq m_{H_{i}^{\pm}} \simeq m_{H_{i}} \simeq m_{A_{i}}~(i=1,2)$ and weak mixing between $\Delta_1$ and $\Delta_2$. 

In addition to the triplet Higgs fields, we introduce a singlet real scalar field $\chi$ as DM candidate. To ensure the candidate stability, the scalar field $\chi$ is assumed to be odd under $Z_2$ symmetry while all SM fields and the triplet Higgs fields are assumed to be even. The scalar potential relevant to DM $\chi$ is given by
\begin{equation}
 V_{\text{DM}}=m_{\chi}^2\chi^2 + \rho_{\chi}\chi^4 + \lambda_{\Phi}\Phi^{\dagger}\Phi \chi^2 + \lambda_{ij}\chi^2 \rm{Tr}(\Delta_i^{\dagger} \Delta_j)\,,
\label{VDM}
\end{equation}
where $i, j=1, 2$ for the hybrid triplet Higgs portal model and $\lambda_{{ij}}=\lambda$ for the single triplet Higgs portal case. The coupling $\lambda_{\Phi}$ is suppressed by DM direct detection experimental results. With the assumption of $\lambda_{\Phi} \ll \lambda_{ij}$, a pair of dark matter $\chi$ dominantly annihilate into $H_i^{\pm\pm}H_j^{\pm\pm}$, $H_i^{\pm}H_j^{\pm}$, $H_iH_j$ and $A_iA_j$.

\section{Electron and positron flux from galaxy DM halo}
It has been generally postulated that the Milky Way galaxy is located in a larger spherical DM halo. For the DM spatial distribution in the galaxy, we considered a generalized NFW profile~\cite{Navarro:1995iw,Navarro:1996gj}  to describe,
\begin{equation}
\rho_{\text{NFW}}(r)=\rho_s\left( \frac{r_s}{r}\right)^{\gamma} \left(1+\frac{r}{r_s} \right)^{\gamma-3}\,,
\end{equation}
where $\rho_s$ and $r_s$ are typical density and typical radius respectively, whose values can fixed by the DM density at  the location of the Sun (local density) and the total DM mass of Milky Way, and $\gamma$ is an undetermined parameter which mainly affects the DM density around the center of Milky Way. According to the astrophysical observations to the rotation curve of the Milky Way, the local density of DM is roughly at the range $0.2-0.6\ \rm{GeV/cm^3}$ \cite{Salucci:2010qr, Iocco:2011jz, Read:2014qva, Pato:2015dua, Green:2017odb} and the total DM mass contained in 60 kpc is estimated to $M_{60}=4.7 \times 10^{11} M_{\odot}$ \cite{SDSS:2008nmx}. With original NFW profile $\gamma=1$, we can fix the typical density and typical radius to  $\rho_s=0.184\, \rm{GeV/cm^3}$ and $r_s=24.42$ kpc for local DM density $\rho_{\odot}=0.3~ \rm{GeV/cm^3}$ at $r_{\odot}=8.5$ kpc. In our model, the DM particles can self-annihilate into charged leptonic final states and generate electron/positron flux. The propagation of $e^{\pm}$ number density energy spectrum $f\equiv dN/dE$ in Milky Way can be described by the diffusion-loss equation,
\begin{equation}
\frac{\partial f}{\partial t} - \nabla(\mathcal{K}(E, \vec{x}) \nabla f) - \frac{\partial}{\partial E}(b(E, \vec{x})f) = Q(E, \vec{x})\,,
\label{propagation}
\end{equation}
where $\mathcal{K}(E, \vec{x})$ is the diffusion coefficient function, which describes the propagation through the turbulent magnetic fields. With assuming the turbulent magnetic fields to be homogeneous and isotropic, the diffusion coefficient function can be parameterized as $\mathcal{K}(E, \vec{x}) = \mathcal{K}_0 (E/ \rm{GeV})^{\delta}$. $b(E, \vec{x})$ is the energy loss coefficient function, which describes energy loss due to inverse-Compton scattering of cosmic microwave background and starlight photons as well as synchrotron radiation. It can be parameterized as $b(E, \vec{x})= b_0 (E/\rm{GeV})^2$~\cite{Cirelli:2010xx}. The electron/positron source term $Q(E, \vec{x})$ is given as 
\begin{equation}
Q(E, \vec{x})=\frac{1}{2}\left(\frac{\rho(\vec{x})}{M_{\text{DM}}}\right)^2 \langle \sigma v \rangle \frac{dN}{dE}\,,
\label{esource}
\end{equation}
where $\rho(\vec{x})$ is the DM density distribution, $M_{\text{DM}}$ is the mass of DM and $\langle \sigma v \rangle$ is the thermal averaged cross section. Considering a steady-case, $\partial f/ \partial t = 0$, and introducing Green's function formalism, the number density energy spectrum of given point in the Galaxy can be written as
\begin{equation}
f(E, \vec{x})=\int d \vec{x}_s \int dE_s \, G(E, \vec{x}; E_s, \vec{x}_s) Q(E, \vec{x})\,,
\end{equation} 
where the Green's function $G(E, \vec{x}; E_s, \vec{x}_s)$ stands for the probability of injected electron or positron at $x_s$ with the energy $E_s$ to reach a location $x$ with the degraded energy $E$. The Green's function satisfies the following equation,
\begin{equation}
\mathcal{K}(E) \nabla^2 G(E, \vec{x}; E_s, \vec{x}_s) + \frac{\partial}{\partial E}(b(E) G(E, \vec{x}; E_s, \vec{x}_s)) = \delta(E-E_s) \, \delta(\vec{x} - \vec{x}_s)\,.
\label{greenf}
\end{equation}
The equation has general solution without boundary condition~\cite{1959The},
\begin{equation}
G(E, \vec{x}; E_s, \vec{x}_s)=\frac{1}{b(E)(4\pi \lambda^2)^{3/2}}\, \text{exp} \left(-\frac{(\vec{x}-\vec{x}_s)^2}{\lambda^2}\right)\,,
\end{equation}
where the diffusion length $\lambda$ is given by
\begin{equation}
\lambda =\theta(E_s - E) \sqrt{\frac{4\mathcal{K}_0(E^{\delta-1}-E_s^{\delta-1})}{b_0(1-\delta)}}\,.
\end{equation}
We solve Eq.~(\ref{greenf}) numerically in a shape of flat cylinder with height $2L$ in the $z$ direction and radius $R = 20\,\text{kpc}$ in the $r$ direction, with considering the number density energy spectrum $f$ vanished on the surface of cylinder. The comparison of solving with and without boundary condition is studied in Ref.~\cite{Delahaye:2007fr}. With the number density energy spectrum $f(E,\vec{x})$, the electron and positron flux is given as
\begin{equation}
\Phi(E,\vec{x})=\frac{v(E)}{4 \pi}f(E,\vec{x})\,,
\end{equation}
where $v(E)$ is the velocity of electron and positron with energy $E$ and is set as $c$ because of $E \gg m_e$.

\section{Origin of Cosmic Ray Positrons Excess at AMS-02}
According to the latest analysis about cosmic ray positrons collected by the Alpha Magnetic Spectrometer (AMS-02)~\cite{Aguilar:2019owu} on the International Space Station, a significant excess starting from $25\, \text{GeV}$ to $1\,\text{TeV}$ is reported, which possibly originates from DM particles annihilation. We choose the same background formula as the latest report by AMS Collaboration \cite{Aguilar:2019owu},
\begin{equation}
\Phi_{e^+}^{\text{bkg}}(E)=\frac{E^2}{\hat{E}^2}\left[ C_d\left( \frac{\hat{E}}{E_1} \right)^{\gamma_d}+ C_s \left( \frac{\hat{E}}{E_2} \right)^{\gamma_s} \text{exp}\left(-\frac{\hat{E}}{E_s} \right) \right]\,,
\end{equation}
where $\hat{E}=E+\varphi_{e^+}$, with the effective solar potential $\varphi_{e^+}$ accounting for the solar effects. We implement the model in {\tt FeynRules} \cite{Alloul:2013bka} to generate the model files and put it in  {\tt micrOMEGAs\,5.0.8} \cite{Belanger:2013oya} to generate the positron flux from DM annihilation. For the propagation of flux, we set the diffusion coefficient $\mathcal{K}_0=0.0112\, \text{kpc}^2/\text{Myr}$, $\delta=0.7$ and the energy loss coefficient $b_0=10^{-16}\, \text{s}^{-1}$.

With the DM mass ($m_{\text{DM}}$) and triplet Higgs mass ($m_{\Delta}$) be fixed, the DM thermal relic density only depend on the coupling $\lambda_{ij}$ in the Eq.~(\ref{VDM}). In other words, the couplings $\lambda_{ij}$ can be fixed by the observed DM thermal relic of $\Omega h^2=0.12$ ~\cite{Planck:2018vyg} and then the thermal averaged cross sections in the Eq.~(\ref{esource}) is also fixed. However, the thermal averaged cross sections obtained from observed DM thermal relic $\langle\sigma v \rangle_0$ is usually much smaller than the required value $\langle\sigma v \rangle$ to account for the excess. In order to fit the AMS-02 data, we need enhancement to the DM annihilation cross sections. The enhancement is usually parameterized by a so-called boost factor (BF) $\kappa_{\text{BF}}=\langle\sigma v\rangle / \langle\sigma v \rangle_0$. The boost factor may originate from small-scale inhomogeneities of the DM density distribution \cite{Kuhlen:2012ft, Dev:2013hka} or Sommerfeld/Breit-Wigner enhancement of DM annihilation~\cite{,Feng:2010zp, Slatyer:2009vg, Cassel:2009wt, Ding:2021zzg, Ibe:2008ye, MarchRussell:2008tu, Guo:2009aj}.

To describe the goodness of the fit, we introduce the chi-square function $\chi^2$, which defined as
\begin{equation}
\chi^2 = \sum_i \frac{\left(f^{\text{th}}_i - f^{\text{exp}}_i \right)^2}{\delta f_i^2}\,,
\end{equation}
where $f^{\text{th}}_i$ are the theoretical predictions, $ f^{\text{exp}}_i$ are the central values of the experimental data and $\delta f_i$ are combined errors of theoretical and experimental data. For the fit to AMS-02 data, we have only considered the experimental errors and used all $E > 1 \text{GeV}$ AMS-02 data (71 data points).
\begin{table}[!t]
\begin{center}
\begin{tabular}{|c| c|c| c| c|c|c| c|c| c|} \hline
\multirow{2}{2cm}{~ Hierarchy~} & \multicolumn{6}{|c|}{$H^{\pm\pm}$ decays} & \multicolumn{3}{|c|}{$H^{\pm}$ decays} \\ [0.5ex] \cline{2-10}
 &$ee$ & $e\mu$ & $e\tau$ & $\mu \mu$ & $\mu \tau$ & ~$\tau \tau$  & $e\nu$ & $\mu \nu$ & $\tau \nu$ \\ [0.5ex] \hline
 NH& 0.49\% & 0.67\% & 4.36\% & 37.37\%  & 34.90\% & 22.21\%& 3.00\% & 55.14\% & 41.86\%  \\ \hline
 IH& 47.53\%  & 1.32\% & 0.84\% & 9.84\%   & 23.74\% & 16.73\%& 49.27\%  & 50.72\% & 0\% \\ \hline
  DH& 25.26\%  & 0.32\% & 0.28\% & 37.83\% & 1.88\% & 34.43\% &24.79\%  & 25.51\% & 49.70\% \\ \hline
 \end{tabular}
\end{center}
\caption{ Branching ratios of charged Higgs leptonic decays for different neutrino mass hierarchies.}
\label{Lbranch}
\end{table} 
In our model, a pair of DM can annihilate to a pair of singly charged triplet Higgs $H^{\pm}$ or doubly charged triplet Higgs $H^{\pm\pm}$, and these particles will further decay into two or four leptonic final states. The Yukawa coupling matrices are fixed by the neutrino mass and oscillation data and the matrix elements are given by the Eq.~(\ref{YNH}--\ref{YDH}) for different neutrino mass hierarchies. As shown in the Table~\ref{Lbranch}, the doubly charged Higgs $H^{\pm\pm}$ dominantly decays into $\mu$ and $\tau$ leptons for the normal neutrino mass hierarchy, while for the inverted hierarchy it dominantly decays to electrons, and for the quasi-degenerate case, the three flavors are almost democratic. When the difference value of DM mass and triplets mass is small ($m_{\text{DM}} \simeq m_{\Delta}$), the leptons generated from triplet Higgs decays have a roughly monochromatic energy spectrum. When $m_{\text{DM}} \gg m_{\Delta}$,  since the triplet Higgses are highly boosted, the leptons will have a continuous energy spectrum.The generated muons and taus will further decay into positron and contribute to the cosmic ray positron flux. The energy spectrum of the positron directly produced by the triplet Higgs is sharp and drop around energy $E=m_{\text{DM}}/2$, while for the positron generated from the muon- and tau-cascade decays, the energy spectrum will be softer and the peak is lower.

\begin{figure}[ht]
  \begin{minipage}{0.48\linewidth}
    \centerline{\includegraphics[width=1\textwidth]{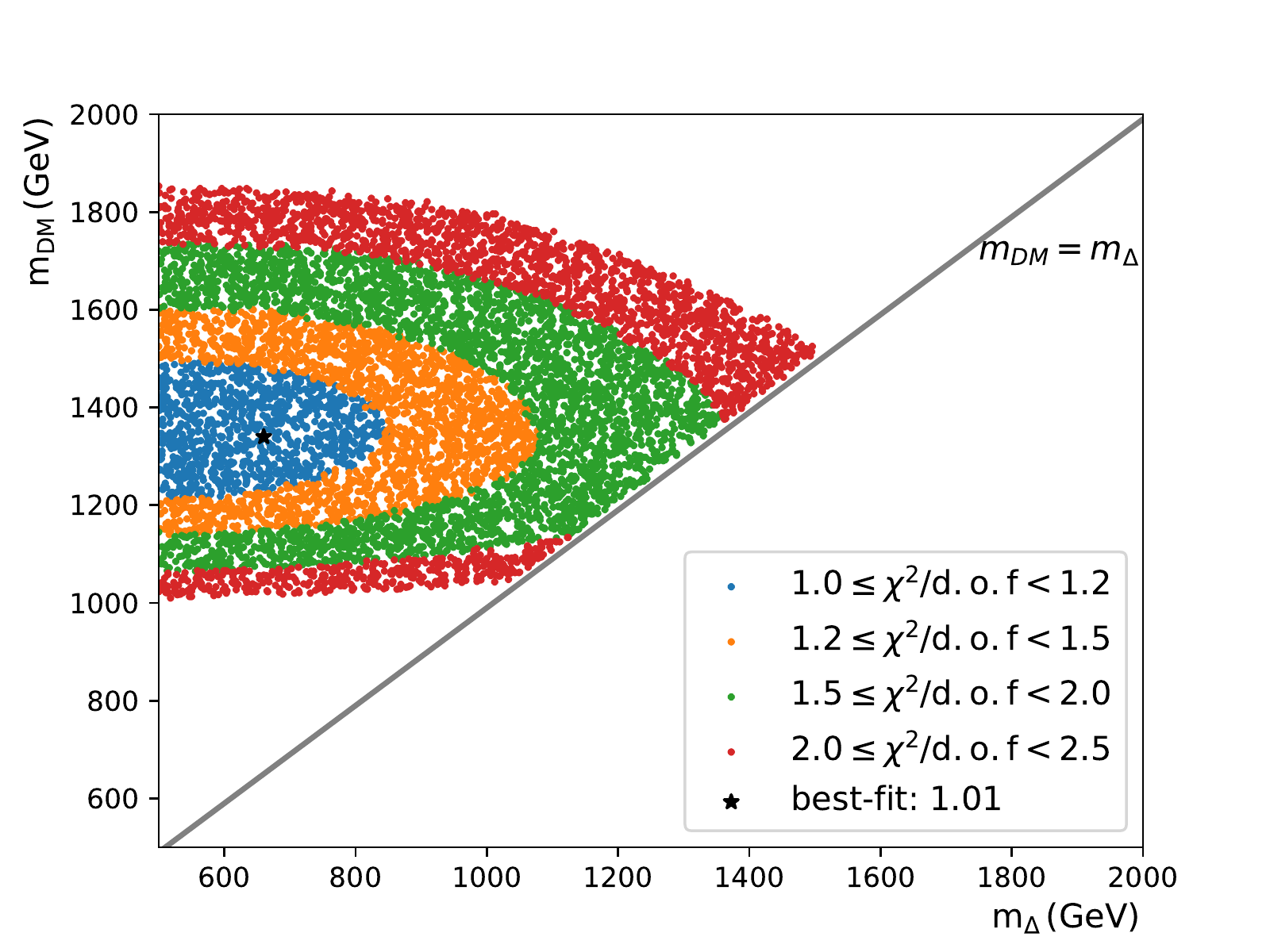}}
    \centerline{$(a)$\ NH case}
  \end{minipage}
  \begin{minipage}{0.48\linewidth}
    \centerline{\includegraphics[width=1\textwidth]{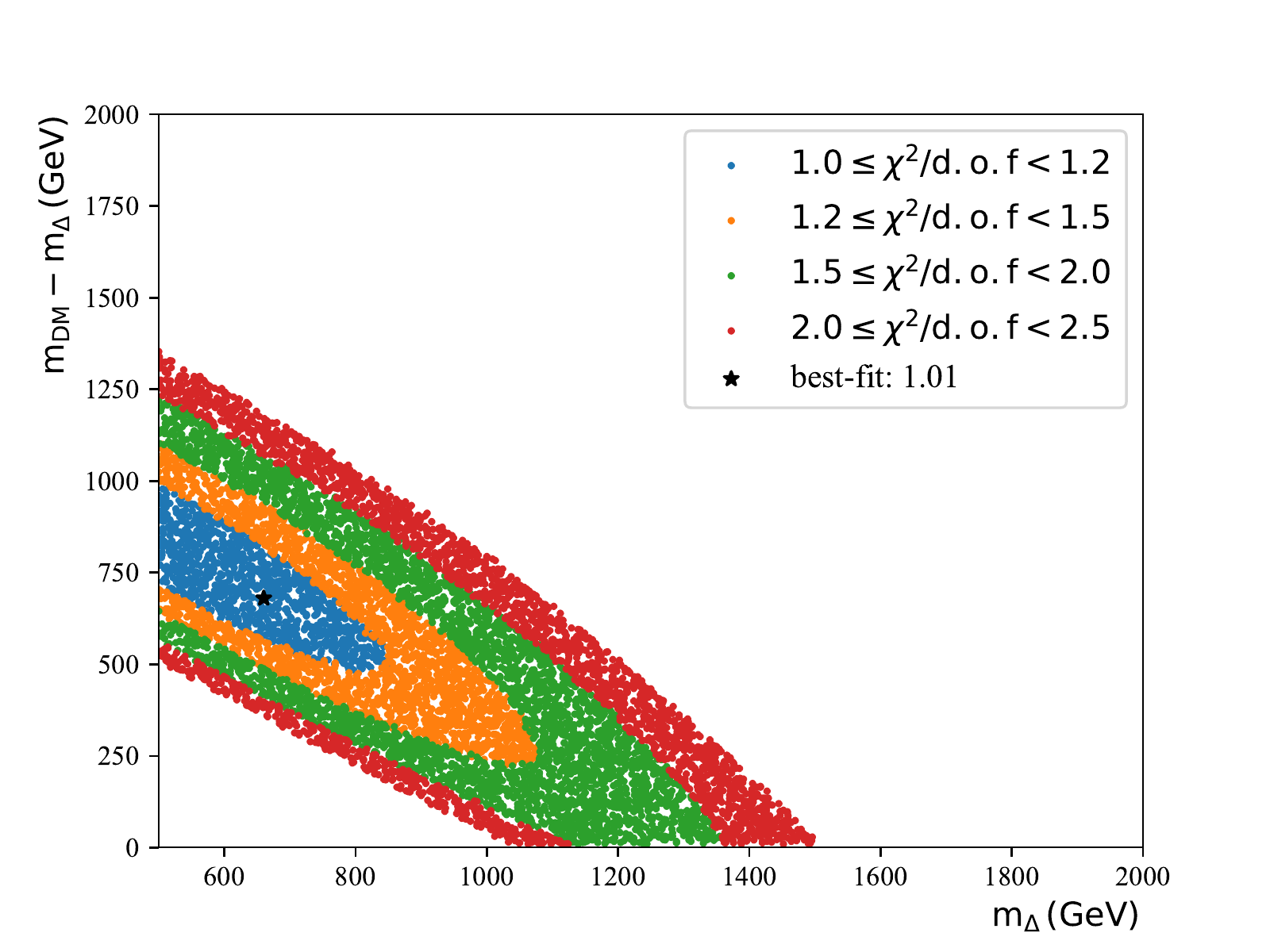}}
    \centerline{$(b)$\ NH case}
  \end{minipage}
   \begin{minipage}{0.48\linewidth}
    \centerline{\includegraphics[width=1\textwidth]{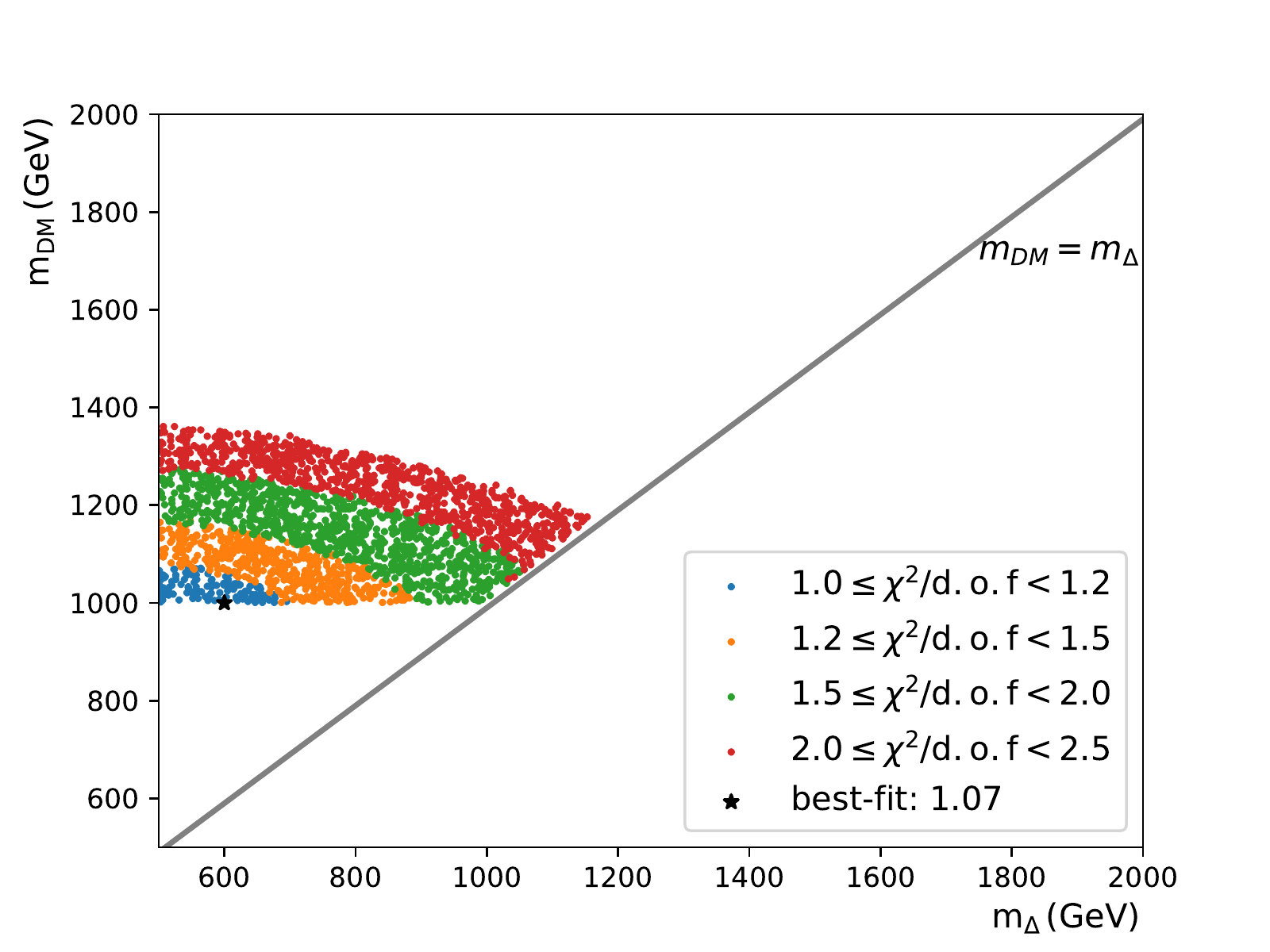}}
    \centerline{$(c)$\ DH case}
  \end{minipage}
  \begin{minipage}{0.48\linewidth}
    \centerline{\includegraphics[width=1\textwidth]{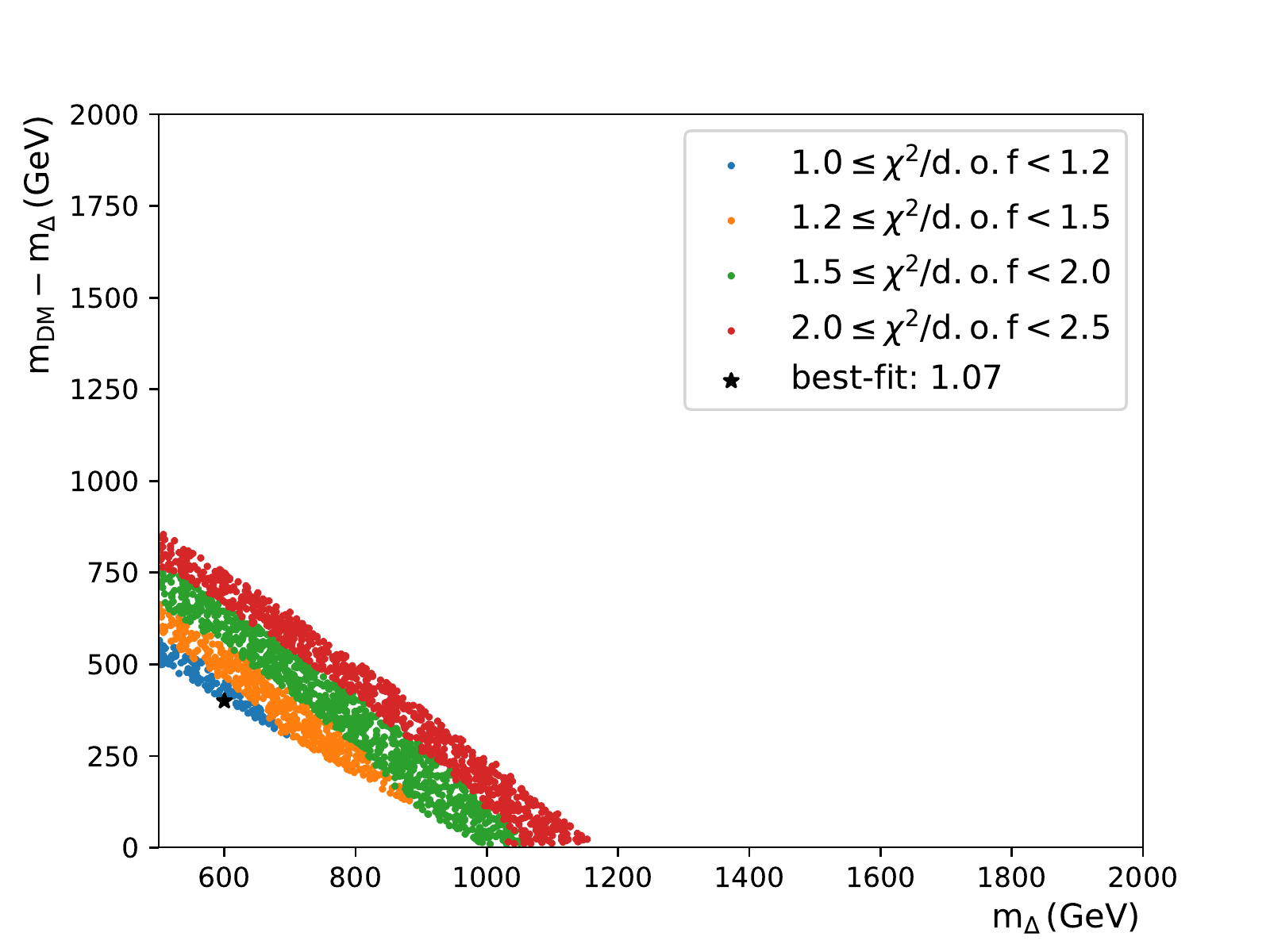}}
    \centerline{$(d)$\ DH case}
  \end{minipage}
  \caption{Favored region of $\chi^2/\text{d.o.f.}$ for the single triplet portal case. The upper and lower panel are dedicated to the normal and degenerate neutrino mass hierarchies respectively. }
  \label{1trip}
\end{figure}

For the single triplet Higgs portal model, we scan over the mass range from 500 GeV to 2 TeV for both DM and triplet Higgs and assume the triplet Higgs mass splitting is negligible ($m_{\Delta} \simeq m_{H} \simeq m_{A}\simeq m_{H^{\pm}} \simeq m_{H^{\pm\pm}}$). As shown in Fig.~\ref{1trip}(a), we present the favored region for $\chi^2/\text{d.o.f.}$  in the plane of the DM mass $m_{\text{DM}}$ versus the triplet Higgs mass $m_{\Delta}$. In the right panel of Fig.~\ref{1trip}(b), we plot the mass difference $(m_{\text{DM}}-m_{\Delta})$ against the triplet Higgs mass for NH scenario for different $\chi^2/\text{d.o.f.}$ values. The black data points in both Fig.~\ref{1trip}(a) and (b) are corresponding to the best fit data points obtained from $\chi^2/\text{d.o.f.}$ analysis for the NH case. Similar plots are generated for DH case and depicted in Fig.~\ref{1trip}(c-d) along with the best fit data points. A comparison between NH and DH scenario reveals that the NH case has a larger range of DM mass about (1-1.8) TeV, for the chosen range of  $\chi^2/\text{d.o.f.}\in (1.0-2.5)$, the corresponding range is restricted into  $(1-1.4)$ TeV for the DH case. Fig.~\ref{1trip}(b) and~\ref{1trip}(d) show that the mass splitting between DM mass and the triplet Higgs in the case of NH is larger than the DH case. Various parameters for the best fit to AMS-02 data with single triplet Higgs portal DM model are tabulated in Table~\ref{1trip_fit} along with the contribution ratios from different DM annihilation channels. It is to be noted for the IH case, $\chi^2/\text{d.o.f.}$ is always found to be larger than 5  for the chosen range of DM  and triplet Higgs mass. Therefore, we don't present the results for the IH scenario.

\begin{figure}[ht!]
  \begin{minipage}{0.48\linewidth}
    \centerline{\includegraphics[width=1\textwidth]{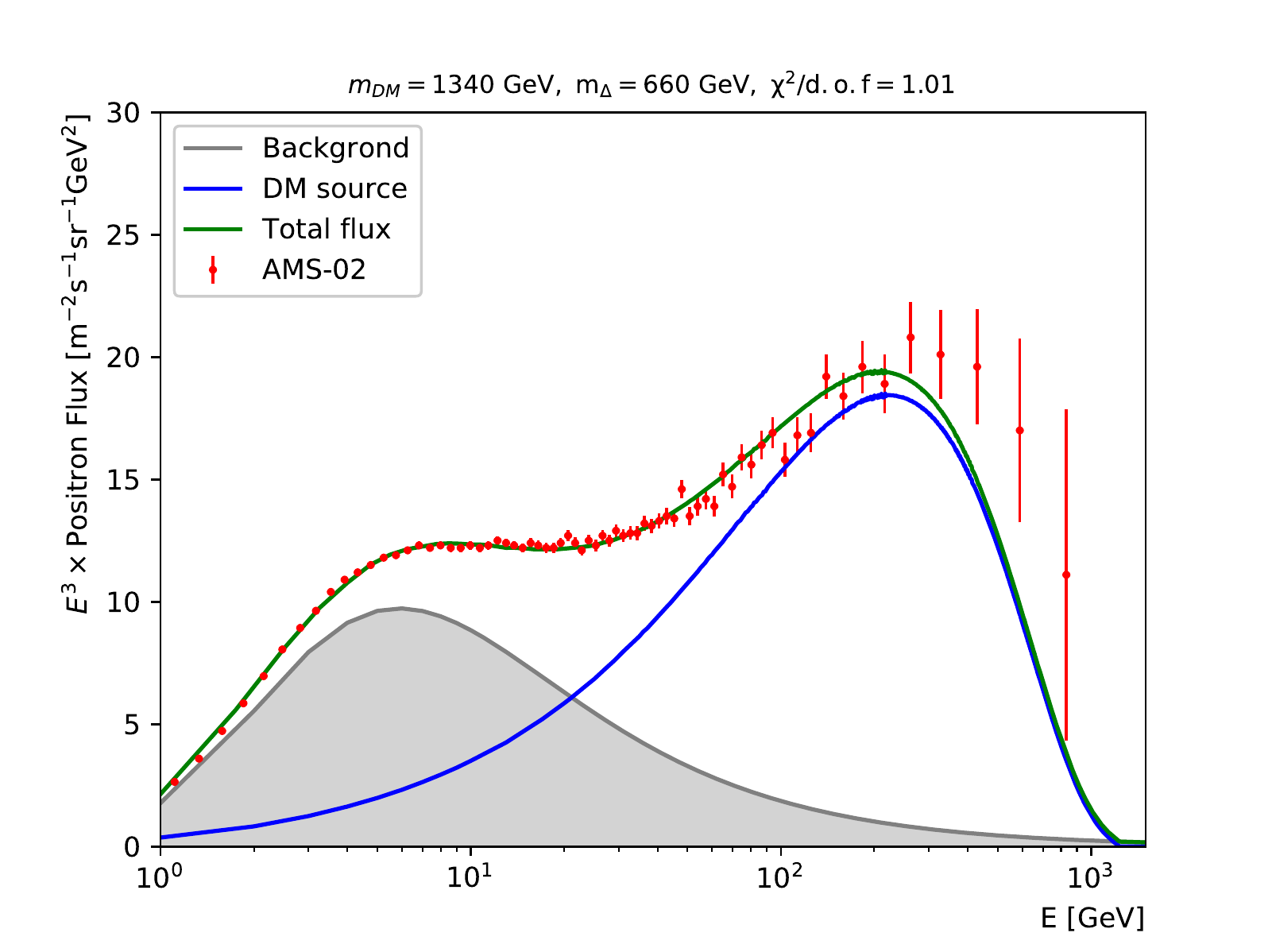}}
    \centerline{$(a)$\ NH case}
  \end{minipage}
  \begin{minipage}{0.48\linewidth}
    \centerline{\includegraphics[width=1\textwidth]{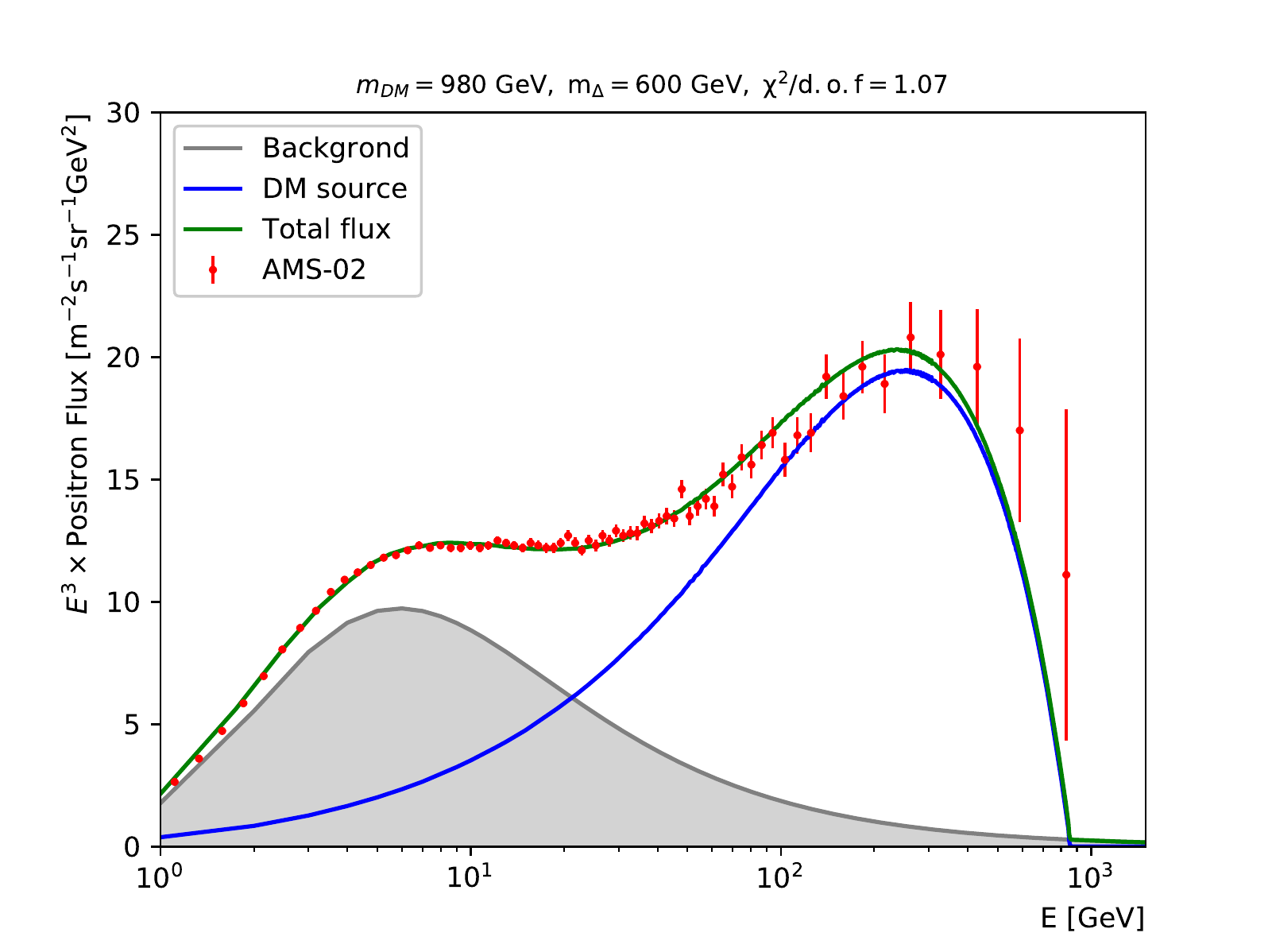}}
    \centerline{$(b)$\ DH case}
  \end{minipage}
  \caption{The best-fit of the positron flux observed by the AMS-02 in the single triplet portal DM model.}
  \label{bestfit1}
\end{figure}

\begin{table}[ht!]
\centering
 \begin{tabular}{|c|c|c|c|c|c|c|c|c|c|c|c|c|}\hline
 Best fit &$m_{\text{DM}}$~(GeV) &  $m_{\Delta}$~(GeV) & $\rm{Br}(H_1 H_1/A_1 A_1)$  & $\rm{Br}(H^+_1 H^-_1/H^{++}_1 H^{--}_1)$  & $\lambda_{\Delta}$ & BF  \\ \hline
  NH & 1340 & 660 & $16\,\%$  & $34\, \%$   & 0.18 & 1758       \\ \hline 
 DH  & 980  & 600 & $17\,\%$  & $33\, \%$   & 0.12 & 608     \\ \hline
 \end{tabular}
  \caption{The best fit data points to AMS-02 positron excess in the single triplet DM model.}
 \label{1trip_fit}
\end{table}

In Fig.~\ref{bestfit1}, we plot positron flux spectrum using the best fit data points obtained for the NH and DH cases (Table~\ref{1trip_fit}). One can notice that the best fit data does not provide a good fit to the positron flux tail at high energy end. The reason is due to that in the single triplet Higgs portal DM model, the flavour structure of the leptonic final states generated by the DM annihilation is fixed by the neutrino oscillation data. The positron flux spectrum range from 10 GeV to 300 GeV indicates that the positrons are dominantly generated by the muon- and/or tau-cascade decays, but not for the case when the energy is beyond 300 GeV. This is also the origin that there are more favored range for the NH case compared with the IH case. In the IH case,  the charged triplet Higgses dominantly decay to electrons (as demonstrated in Table~\ref{Lbranch}). In the following, we turn to study of the hybrid triplet Higgs portal DM and we find that the fit getting much better, especially for the IH case. 
\begin{figure}[ht]
  \begin{minipage}{0.3\linewidth}
    \centerline{\includegraphics[width=1\textwidth]{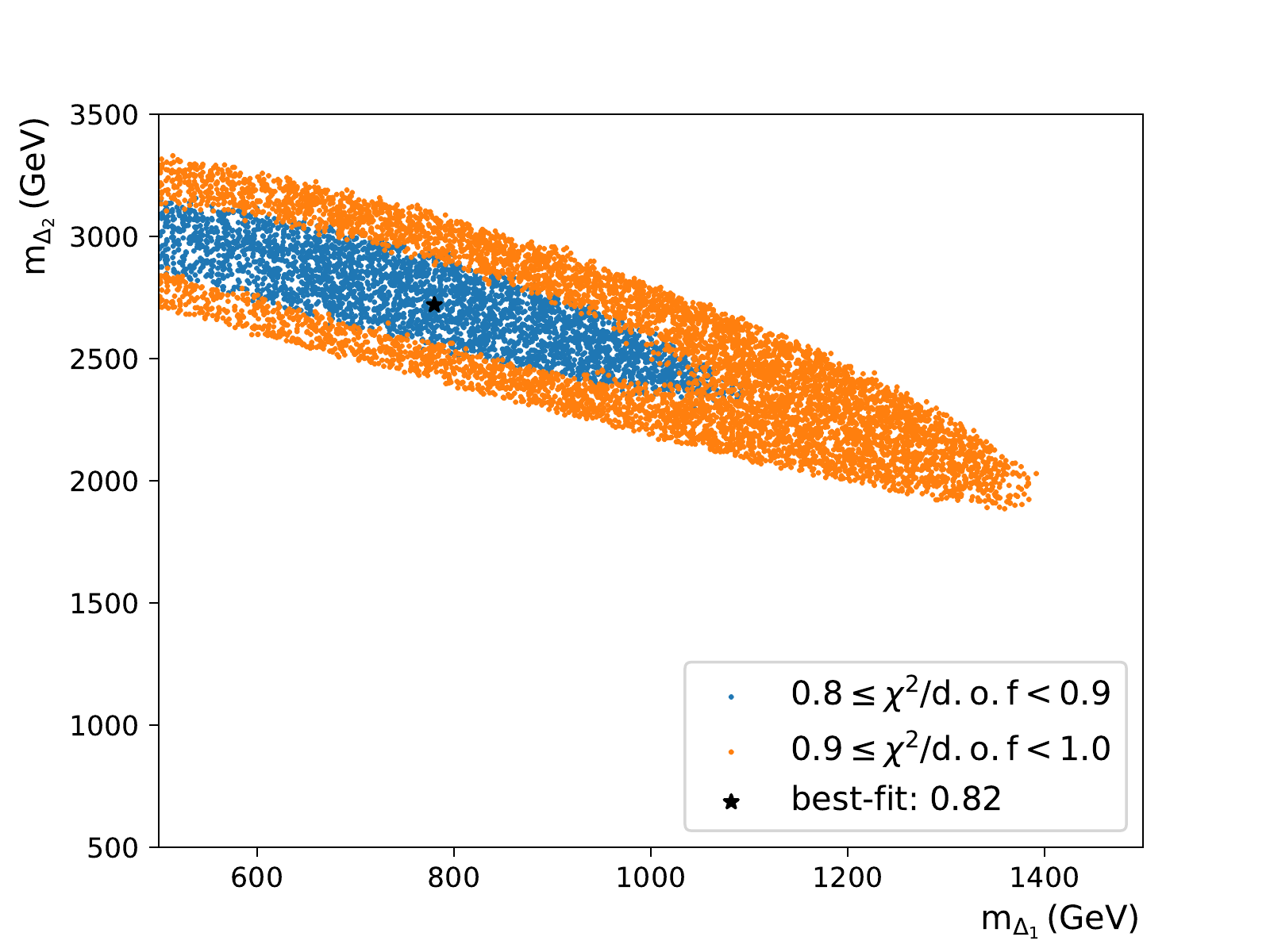}}
    \centerline{$(a)$\ NH case}
  \end{minipage}
  \begin{minipage}{0.3\linewidth}
    \centerline{\includegraphics[width=1\textwidth]{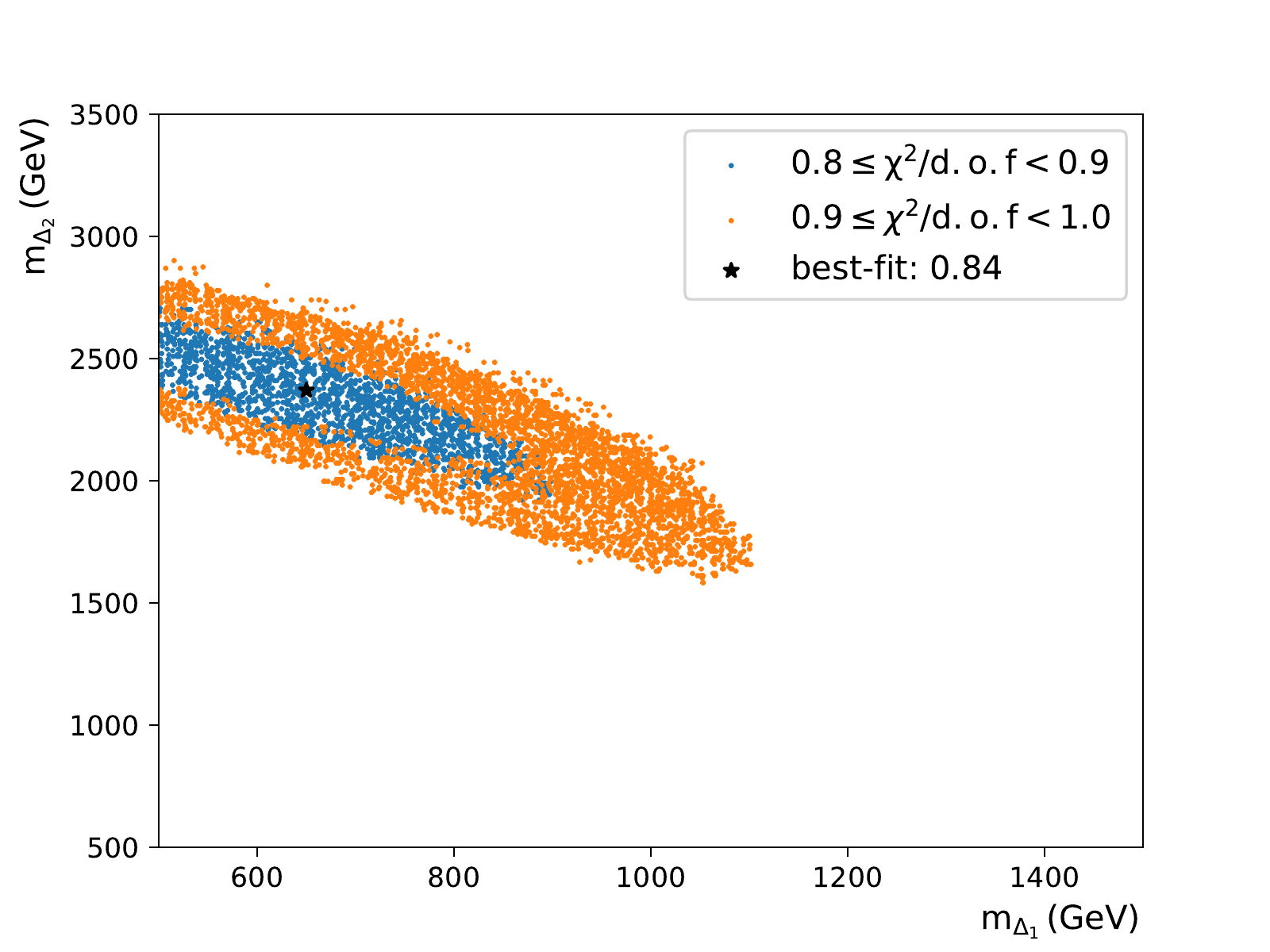}}
    \centerline{$(b)$\ IH case}
  \end{minipage}
  \begin{minipage}{0.3\linewidth}
    \centerline{\includegraphics[width=1\textwidth]{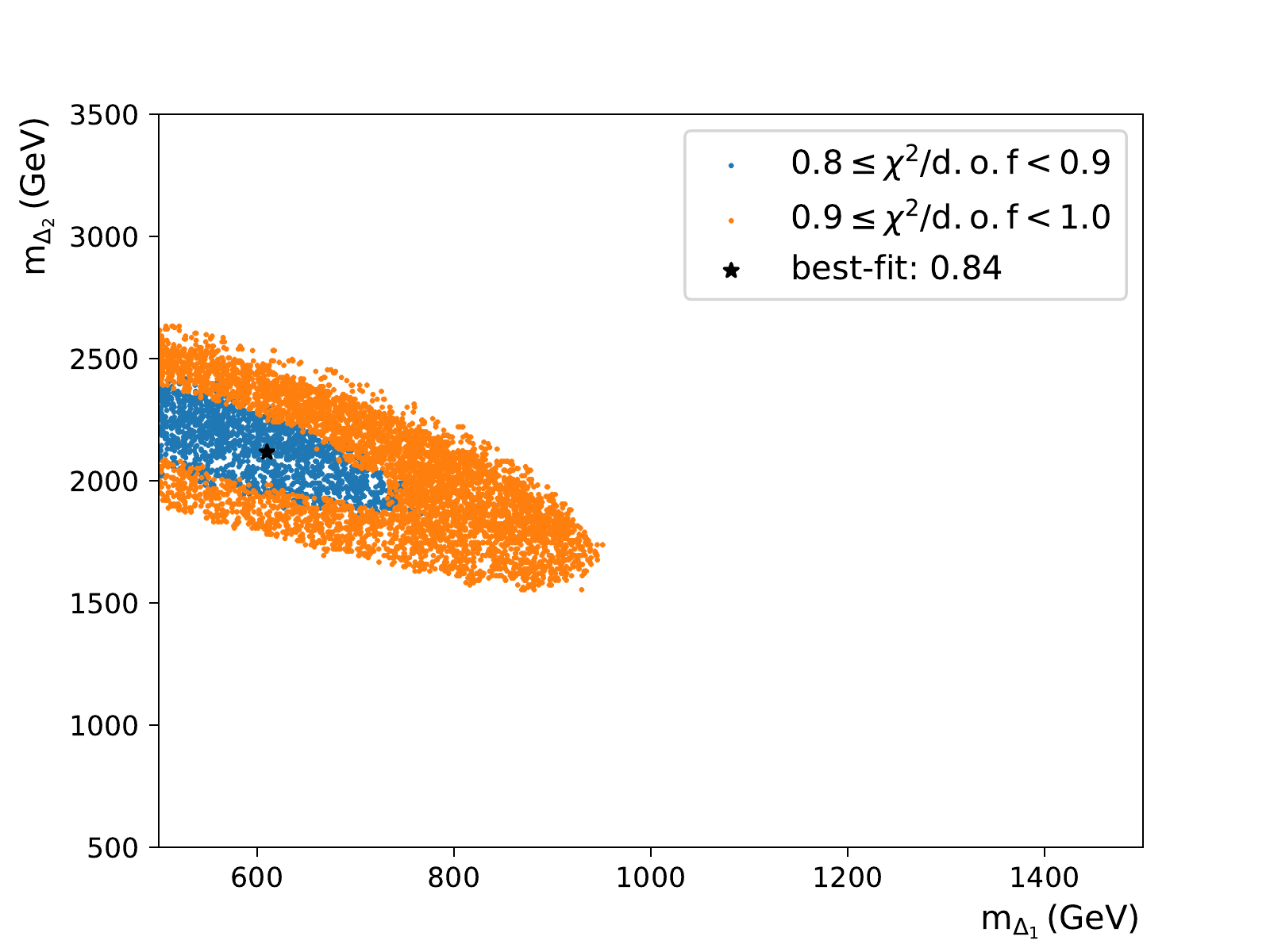}}
    \centerline{$(c)$\ DH case}
  \end{minipage}
  \caption{Favored region with $\chi^2/\text{d.o.f.}$ for the hybrid triplet portal DM, for the case of the mass difference taken as $2m_{\text{DM}}-(m_{\Delta_1}+m_{\Delta_2})= 10\ \text{GeV}$.}
  \label{2trip}
\end{figure}

\begin{table}[h!]
\centering
\resizebox{16cm}{1.1cm}{
 \begin{tabular}{|c|c|c|c|c|c|c|c|c|c|c|c|c|}\hline
 Best fit &$m_{\Delta1}$~(GeV)&  $m_{\Delta2}$~(GeV) & $\rm{Br}(H_1 H_1/A_1 A_1)$ & $\rm{Br}(H^+_1 H^-_1/H^{++}_1   H^{--}_1)$ & $\rm{Br}(H_1 H_2/A_1 A_2)$ & $\rm{Br}(H^+_1 H^-_2/H^{++}_1 H^{--}_2)$  & $\lambda_{11}$ &  $\lambda_{12}$ & BF \\ \hline
  NH & 780 & 2720 & $12\,\%$  & $25\, \%$ & $4\, \%$     &  $9\, \%$    & 0.10 & 0.10 & 2100        \\ \hline 
  IH & 650 & 2370 & $15\,\%$  & $30\, \%$ & $2\, \%$     &  $3\, \%$    & 0.14 & 0.09 & 1126    \\ \hline 
  DH & 590 & 2210 & $13\,\%$  & $29\, \%$ & $3\, \%$     &  $5\, \%$    & 0.12 & 0.10 & 1267  \\ \hline
 \end{tabular}}
 \caption{Best fit data points to AMS-02 positron excess for hybrid triplet Higgs portal DM model. The ``Branch ratios'' indicate the branch ratios of the DM annihilation channels. ``BF'' denote the needed enhancement factors.}
 \label{2trip_fit}
\end{table}

We now fit the AMS-02 data by the hybrid triplet portal DM, with a light triplet Higgs and
a heavy triplet Higgs final states as the DM main annihilation channels. We find the positron flux can be best fit when the light triplet Higgs mainly decay to muons and taus whereas the heavy one mainly decay to electrons. In order to generate the high energy region positrons by the heavier triplet Higgs decays, we open $\chi \chi \to \Delta_1 \Delta_1$ and $\chi \chi \to \Delta_1 \Delta_2$ two annihilation channels, while the channel $\chi \chi \to \Delta_2 \Delta_2$ is forbidden by setting $m_{\Delta_1}+m_{\Delta_2} < 2m_{\text{DM}} < 2m_{\Delta_2}$. Without loss of generality, as a benchmark, we set $2m_{\text{DM}}-(m_{\Delta_1}+m_{\Delta_2})= 10$ GeV. We scan over the light triplet Higgs mass $m_{\Delta_1}$ ranging from 500 GeV to 1.5 TeV while the heavier triplet Higgs mass is varied within the range from 1.5 TeV to 3.5 TeV. In Fig.~\ref{2trip} (a-c), we plot the allowed range of $m_{\Delta_1}$ versus $m_{\Delta_2}$ parameter space which fit the AMS-02 data with $0.8\leq \chi^2/\text{d.o.f.} \leq 1$ for the NH, IH and DH scenarios respectively.

\begin{figure}[ht]
  \begin{minipage}{0.3\linewidth}
    \centerline{\includegraphics[width=1\textwidth]{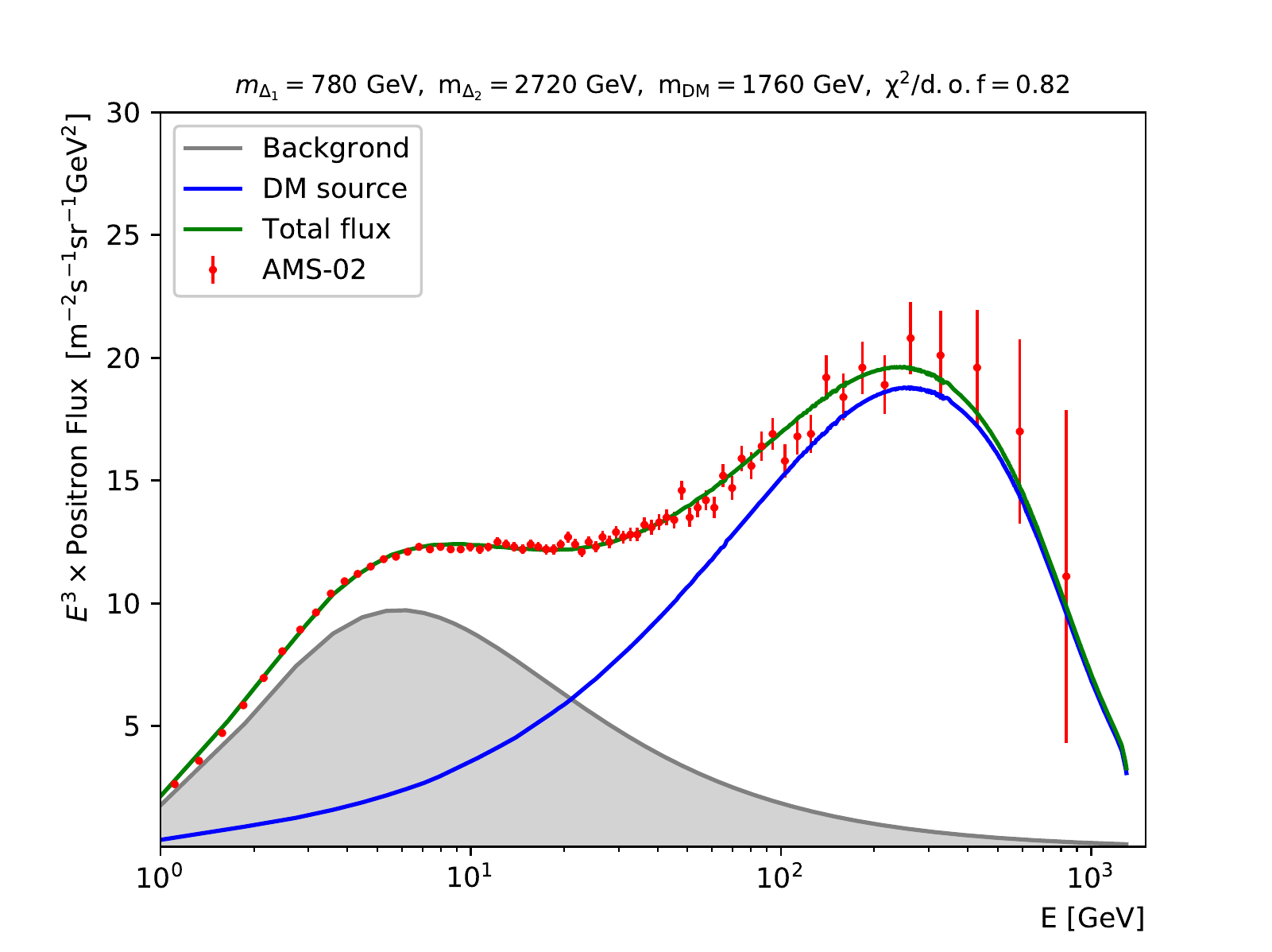}}
    \centerline{$(a)$\ NH case}
  \end{minipage}
  \begin{minipage}{0.3\linewidth}
    \centerline{\includegraphics[width=1\textwidth]{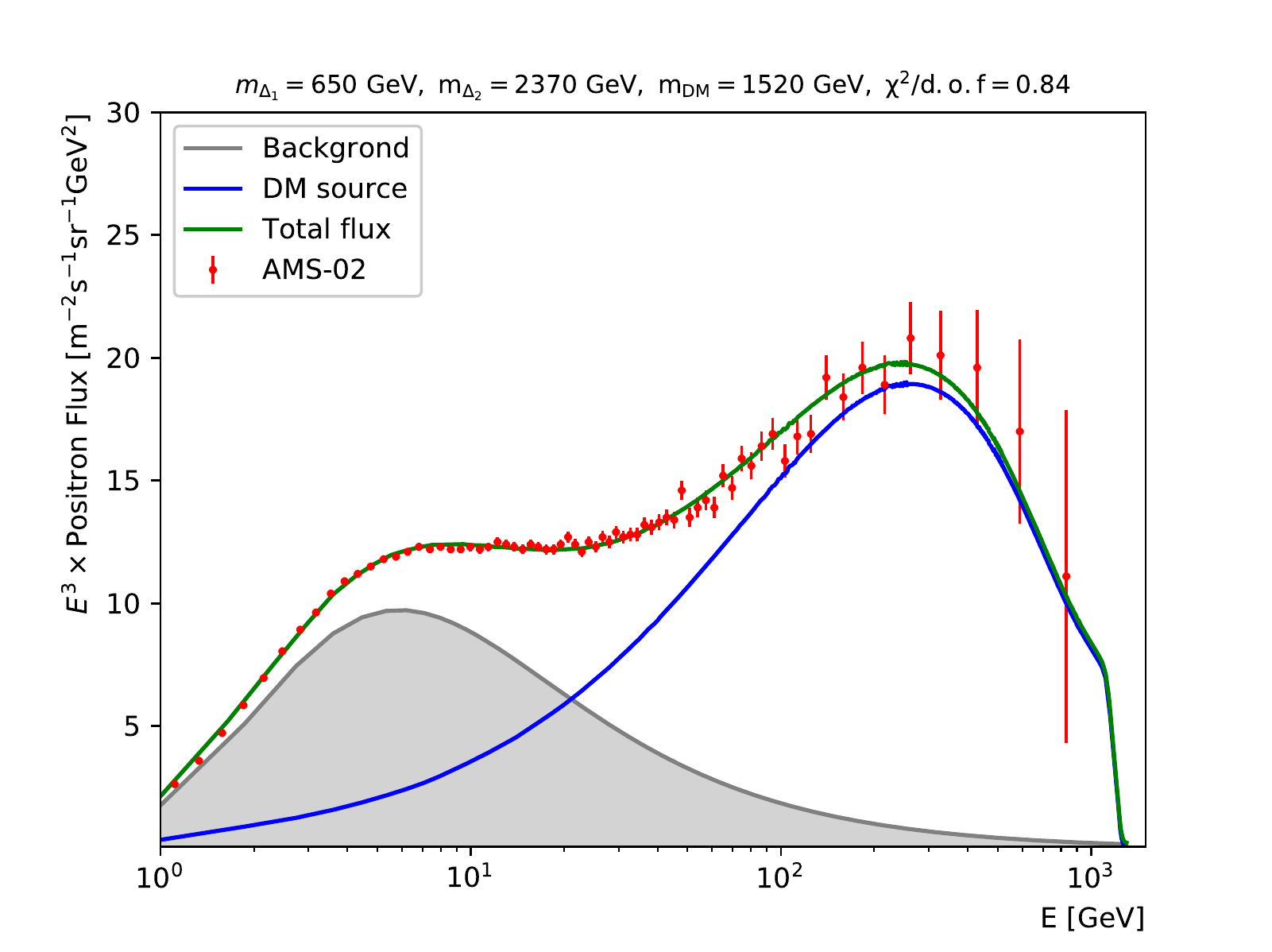}}
    \centerline{$(b)$\ IH case}
  \end{minipage}
  \begin{minipage}{0.3\linewidth}
    \centerline{\includegraphics[width=1\textwidth]{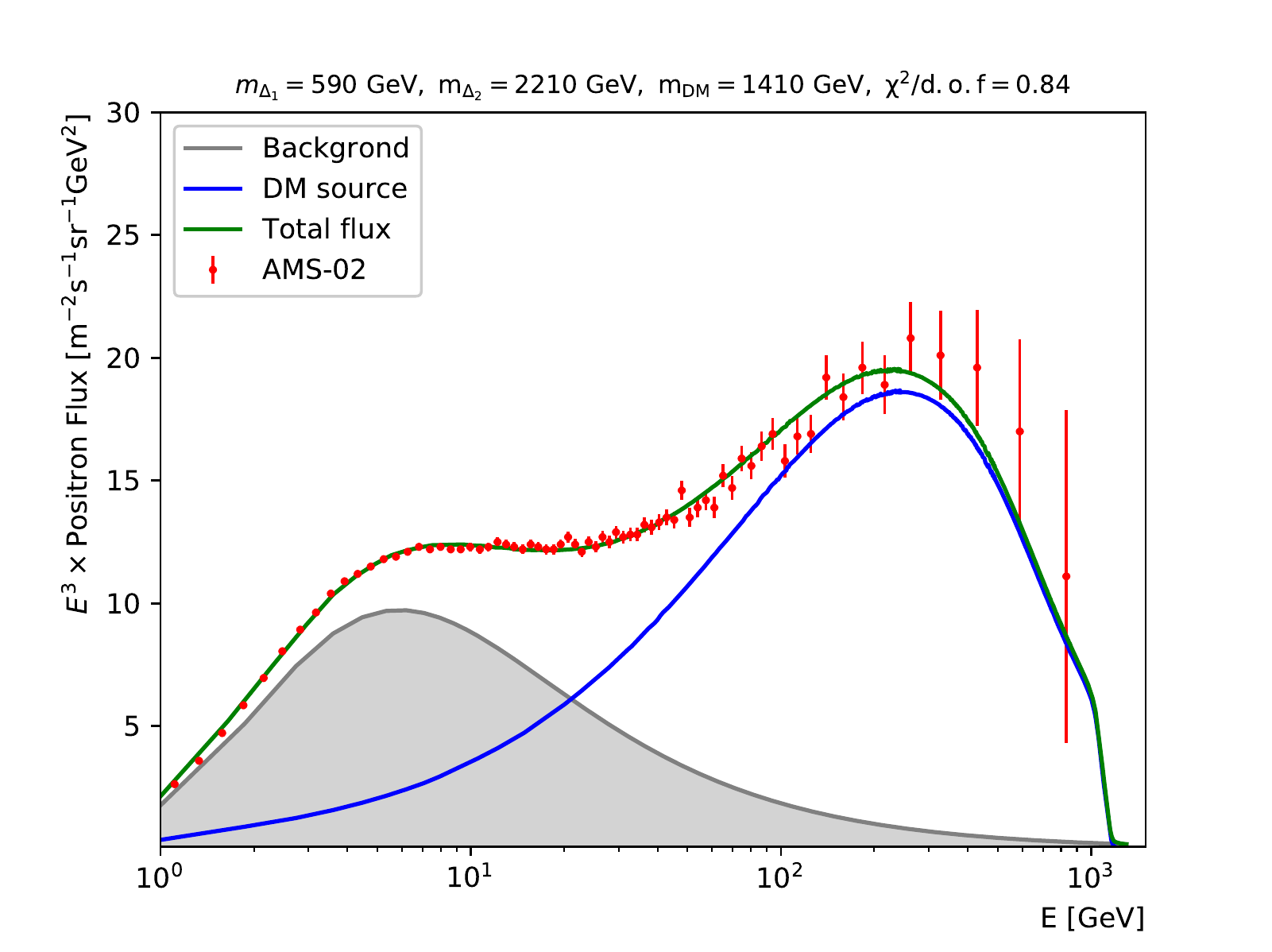}}
    \centerline{$(c)$\ DH case}
  \end{minipage}
  \caption{The best fit to the positron flux observed by the AMS-02 in the hybrid triplet portal DM model.}
  \label{bestfit2}
\end{figure}

The favored region for the masses of DM and the triplet Higgses are shown in Fig.~\ref{2trip} and the best fit data sets are given in Table~\ref{2trip_fit}, with the corresponding positron flux results shown in Fig.~\ref{bestfit2}. One intriguing feature of the hybrid triplet portal DM is that it can provide a pretty good fit to the AMS-02 data for the IH case, unlike the single triplet portal DM model discussed earlier. Apart from the above, we also find that the hybrid triplet portal DM can provide better fit for the excess at high energy region for the energy above 300 GeV. This unique feature of the hybrid triplet model is due to that the heavy triplet Higgses produce hard positrons while the light triplet Higgses generate the soft positrons. Finally, we make a direct comparison between the single triplet Higgs portal DM model and the hybrid triplet Higgs portal DM model. The $\chi^{2}/\text{d.o.f.}$ of the best fit points in varied scenarios are listed in Table~\ref{comparison}, which clearly indicates that the hybrid triplet Higgs portal model provides much better fit to the AMS-02 data. Especially, the fit gets significant improved for the neutrino mass IH scenario.

\begin{table}[!ht]
 \begin{center}
 \begin{tabular}{|c|c|c|c|}
 \hline   & Normal ordering & Inverted ordering & Degenerate ordering \\ \hline
 single triplet portal & 1.01 &  8.12 & 1.07 \\ \hline
hybrid triplet portal & 0.82 &  0.84 & 0.82 \\ \hline
 \end{tabular}
 \end{center}
 \caption{The $\chi^{2}/\text{d.o.f.}$ of the best fit for different neutrino mass hierarchies in the single and hybrid triplet portal DM scenarios.}
 \label{comparison}
\end{table}
Note that the DM annihilation also produce gamma-rays. We compared the gamma-rays produced by the DM annihilation with the isotropic diffuse gamma-ray background (IGRB) observed by the Fermi-LAT experiment~\cite{Fermi-LAT:2014ryh} and we find the predicted gamma-rays are slightly constrained by the Fermi-LAT observation in the energy region beyond hundreds GeV. Earlier work about the single triplet Higgs portal model reached similar conclusion~\cite{Li:2018abw}. A feasible way to release the inconsistency between the DM origin of cosmic positron excesses and the IGRB constraints is to introduce a dark matter disk instead of simple dark matter halo as dark matter density distribution~\cite{Belotsky:2016tja}. More studies on the compatibility between the DM origin of cosmic positron excesses and gamma-ray limits were performed in Ref.~\cite{Cirelli:2012ut, Liu:2016ngs, Xiang:2017jou, Belotsky:2018vyt, Belotsky:2019xti}. 

\section{Origin of Cosmic-Ray Electron/Positron Excesses at DAMPE}
In this section, we attempt to explain the cosmic ray electron and positron excesses observed by the DAMPE satellite. Different from AMS-02, DAMPE doesn't distinguish the charge of electron events and collect the sum of electron and positron events. The DAMPE data has been considered to be the electron and positron flux assisted with a smoothly broken power law spectrum~\cite{Ambrosi:2017wek}. The signals have also been considered as an excess in the range (0.6-1.5) TeV from DM sources with single broken power-law background and double broken power-law background~\cite{Fan:2017sor, Ge:2017tkd, Ge:2020tdh} or excess in the range (0.1-1.5) TeV with single power law background~\cite{Liu:2019iik}. Here we assume that there exists a smooth excess in the energy range (0.1-1.5) TeV along with an excess peak at 1.5 TeV. We consider that the continuous excess of electron and positron flux is attributed to the DM annihilation within the Milky Way (MW) and the sharp peak at 1.5 TeV originates mainly from a nearby DM sub-halo. Meanwhile, we take a simple single power law background to fit the background for $E<100$ GeV and $E>1.5$ TeV and the background formula of electron and positron flux is expressed as
\begin{equation}
\Phi_{e^{\pm}}^{\text{bkg}} = C \left(\frac{E}{\rm{GeV}}\right)^{-\gamma},
\end{equation}
where we set the free parameter: $C=450\ \rm{(GeV\cdot m^2\cdot s\cdot sr)^{-1}}$ and $\gamma = 3.26$ by fitting the first two points and the four points between 1.5 TeV and 2.5 TeV of the DAMPE data.
\begin{figure}[ht]
  \begin{minipage}{0.48\linewidth}
    \centerline{\includegraphics[width=1\textwidth]{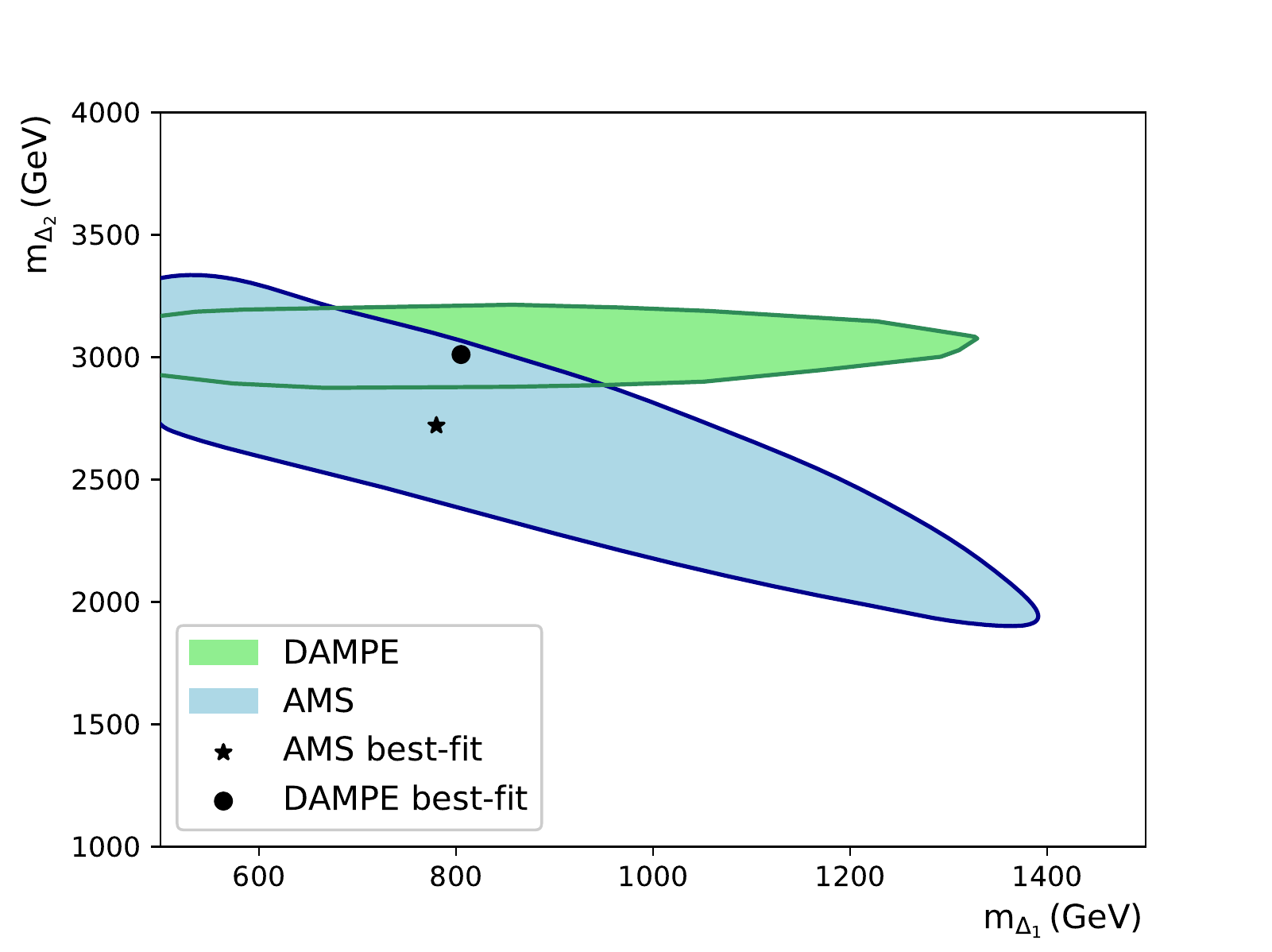}}
    \centerline{(a)}
  \end{minipage}
  \begin{minipage}{0.48\linewidth}
    \centerline{\includegraphics[width=1\textwidth]{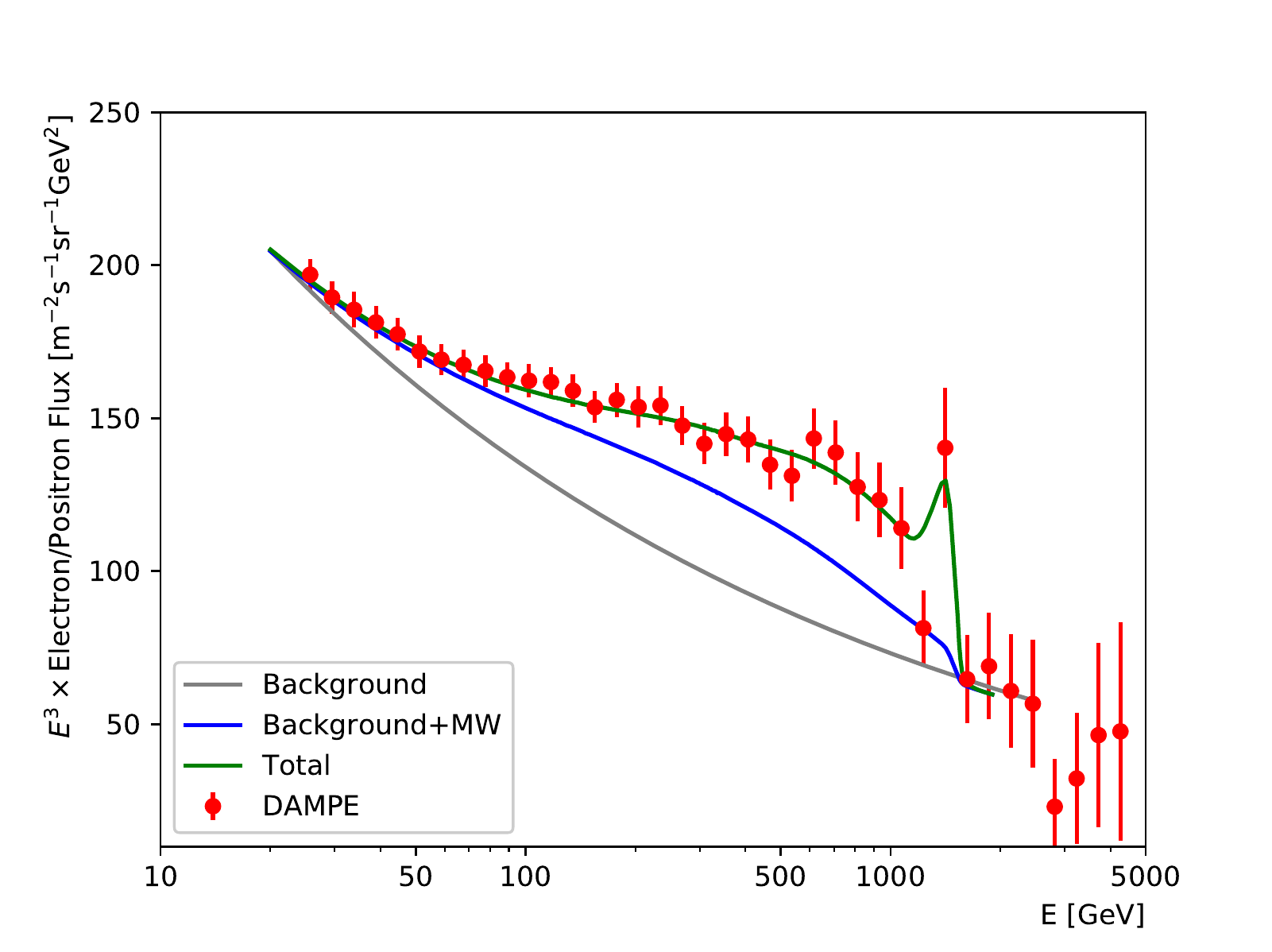}}
    \centerline{(b)}
  \end{minipage}
  \caption{(a): The allowed parameter space of $m_{\Delta_1}$ versus $m_{\Delta_2}$ that fit both the AMS-02 data with $0.8 \leq \chi^2/\text{d.o.f.} \leq 1$ and the DAMPE data with  $0.2 \leq \chi^2/\text{d.o.f.} \leq 0.3$ for NH case. (b): The best fit to the DAMPE CRE spectrum with $m_{\Delta_1} = 810\ \rm{GeV}$, $m_{\Delta_2} = 3000\ \rm{GeV}$ and $\chi^2/\text{d.o.f.} = 0.2$. The blue line indicates the contribution from DM annihilation in the Milky Way DM halo and the green line includes the contribution of the nearby sub-halo.} 
  \label{bestfitDAMPE}
\end{figure}

To explain the peak around 1.5 TeV in the DAMPE data, the heavy triplet Higgs mass $m_{\Delta_2}$ is required to be set around 3 TeV and the heavy charged triplet Higgs mainly decays to electron flavor. From the results of fitting AMS-02 data, we learn that the heavy triplet Higgs can be chosen as high as 3 TeV only in the normal neutrino mass hierarchy scenario, therefore we choose the NH case to fit the DAMPE electron and position excess in the hybrid triplet Higgs portal model. We scan over the light triplet Higgs mass $m_{\Delta_1}$ ranging from 500 GeV to 1.5 TeV and the heavy triplet Higgs mass $m_{\Delta_2}$ ranging from 1.5 TeV to 4 TeV.  We obtain the allowed range of $m_{\Delta_1}$ versus $m_{\Delta_2}$ that fit both the AMS-02 data with $0.8 \leq \chi^2/\text{d.o.f.} \leq 1$ and the DAMPE data with  $0.2 \leq \chi^2/\text{d.o.f.} \leq 0.3$ for NH case, as shown in Fig.~\ref{bestfitDAMPE}. For the dark matter sub-halo profile, we adopt the NFW density profile with $\rho_s = 100\ \rm{GeV/cm^3}$, $r_s = 0.1$ kpc and $\gamma = 0.5$ and the distance of between the Sun and the center of the sub-halo is set as $d_s = 0.2$ kpc.

\section{Summary}
In the work, we study a hybrid triplet Higgs portal DM model by extending the SM with two triplet scalar fields and a singlet scalar DM, which can generate neutrino mass by a hybrid Type-II seesaw mechanism and provide DM candidate. In this framework, the DM particles can self-annihilate into triplet Higgses which further cascade decay into leptons. We calculate the energy spectrum of the electron and positron flux from the DM annihilation in the Galactic halo to explain the flavor structure of both recent AMS-02 positron data and DAMPE electron/positron data. We revisit the single triplet Higgs portal DM model to address the AMS-02 positron excess and obtain the favored parameters region for different neutrino mass hierarchy. We then use the hybrid triplet Higgs portal DM to fit the AMS-02 data. We find that the hybrid triplet Higgs portal model can provide much better fit, especially for the inverted hierarchy neutrino mass scenario. We also perform a fit to the DAMPE electron/positron data using the hybrid triplet Higgs portal model and find that the NH scenario has favored region for fitting both the AMS-02 and the DAMPE signals.

\noindent{\bf Acknowledgements.}
This work is supported in part by the National Science Foundation of China (11775093, 12175082).

\begin{appendices}
\section{APPENDIX}
Here we present the model parameter values corresponding to the best fit points (BFPs) in Table~\ref{1trip_fit} and Table~\ref{2trip_fit}.

\noindent{BFP~1: Input parameters for NH case in Table~\ref{1trip_fit}:}
\begin{align*}
&m_{H}, m_{A}, m_{H^{\pm}}, m_{H^{\pm\pm}}=660\,\text{GeV}, \quad m_{\text{DM}}=1340\,\text{GeV},\\
&\lambda_{\Delta}=0.18, \quad \sin \alpha =0,\quad v_{\Delta} =  10^{-6}\, \text{GeV}.
\end{align*}

\noindent{BFP~2: Input parameters for DH case in Table~\ref{1trip_fit}:}
\begin{align*}
&m_{H}, m_{A}, m_{H^{\pm}}, m_{H^{\pm\pm}}=600\,\text{GeV}, \quad m_{\text{DM}}=980\,\text{GeV},\\
&\lambda_{\Delta}=0.12, \quad \sin \alpha =0,\quad v_{\Delta} =  10^{-6}\, \text{GeV}.
\end{align*}

\noindent{BFP~3: Input parameters for NH case in Table~\ref{2trip_fit}:}
\begin{align*}
&m_{12}=-120\,\text{GeV}, \quad \bm{\rho}=\bm{\rho}^{\prime}=\bm{\sigma}=\begin{pmatrix}
0.5 & 0.0001 \\ 0.0001 & 0.5
\end{pmatrix}, \quad \bm{\sigma}^{\prime}=\begin{pmatrix}
0.5 & 0.0001 \\ 0.0001 & 0.5
\end{pmatrix},\\
&\rho_{\Delta}=\rho_{\Delta}^{\prime}=0.1, \quad \bm{\mu}=\begin{pmatrix}
-9.5 \times 10^{-6}\,\text{GeV} \\ -1.2 \times 10^{-4}\,\text{GeV}
\end{pmatrix}, \quad \bm{V}=\begin{pmatrix}
10^{-6} \,\text{GeV} \\ 10^{-6} \,\text{GeV}
\end{pmatrix}, \\
&m_{\chi}=1770\,\text{GeV}, \quad \lambda_{\Phi}=0, \quad \lambda_{11}=0.10, \quad \lambda_{12}=0.10\,.
\end{align*}
With these parameter values, the particle masses are determined as,
\begin{align*}
&m_{H_1}, m_{A_1}, m_{H_1^{\pm}}, m_{H_1^{\pm\pm}}=780\,\text{GeV}, \quad m_{H_2}, m_{A_2}, m_{H_2^{\pm}}, m_{H_2^{\pm\pm}}=2720\,\text{GeV}, \quad m_{\text{DM}}=1760\,\text{GeV}.
\end{align*}

\noindent{BFP~4: Input parameters for IH case in Table~\ref{2trip_fit}:}
\begin{align*}
&m_{12}=-120\,\text{GeV}, \quad \bm{\rho}=\bm{\rho}^{\prime}=\bm{\sigma}=\begin{pmatrix}
0.5 & 0.0001 \\ 0.0001 & 0.5
\end{pmatrix}, \quad \bm{\sigma}^{\prime}=\begin{pmatrix}
0.5 & 0.0001 \\ 0.0001 & 0.5
\end{pmatrix},\\
&\rho_{\Delta}=\rho_{\Delta}^{\prime}=0.1, \quad \bm{\mu}=\begin{pmatrix}
-6.5 \times 10^{-6}\,\text{GeV} \\ -9.2 \times 10^{-5}\,\text{GeV}
\end{pmatrix}, \quad \bm{V}=\begin{pmatrix}
10^{-6}\,\text{GeV} \\ 10^{-6}\,\text{GeV}
\end{pmatrix}, \\
&m_{\chi}=1520\,\text{GeV}, \quad \lambda_{\Phi}=0, \quad \lambda_{11}=0.14, \quad \lambda_{12}=0.09\,.
\end{align*}
With these parameter values, the particle masses are determined as,
\begin{align*}
&m_{H_1}, m_{A_1}, m_{H_1^{\pm}}, m_{H_1^{\pm\pm}}=650\,\text{GeV}, \quad m_{H_2}, m_{A_2}, m_{H_2^{\pm}}, m_{H_2^{\pm\pm}}=2370\,\text{GeV}, \quad m_{\text{DM}}=1520\,\text{GeV}.
\end{align*}

\noindent{BFP~5: Input parameters for DH case in Table~\ref{2trip_fit}:}
\begin{align*}
&m_{12}=-120\,\text{GeV}, \quad \bm{\rho}=\bm{\rho}^{\prime}=\bm{\sigma}=\begin{pmatrix}
0.5 & 0.0001 \\ 0.0001 & 0.5
\end{pmatrix}, \quad \bm{\sigma}^{\prime}=\begin{pmatrix}
0.5 & 0.0001 \\ 0.0001 & 0.5
\end{pmatrix},\\
&\rho_{\Delta}=\rho_{\Delta}^{\prime}=0.1, \quad \bm{\mu}=\begin{pmatrix}
-5.3 \times 10^{-6}\,\text{GeV} \\ -8.0 \times 10^{-5}\,\text{GeV}
\end{pmatrix}, \quad \bm{V}=\begin{pmatrix}
10^{-6}\,\text{GeV} \\ 10^{-6}\,\text{GeV}
\end{pmatrix}, \\
&m_{\chi}=1410\,\text{GeV}, \quad \lambda_{\Phi}=0, \quad \lambda_{11}=0.12, \quad \lambda_{12}=0.10\,.
\end{align*}
With these parameter values, the particle masses are determined as,
\begin{align*}
&m_{H_1}, m_{A_1}, m_{H_1^{\pm}}, m_{H_1^{\pm\pm}}=590\,\text{GeV}, \quad m_{H_2}, m_{A_2}, m_{H_2^{\pm}}, m_{H_2^{\pm\pm}}=2210\,\text{GeV}, \quad m_{\text{DM}}=1410\,\text{GeV}.
\end{align*}

\end{appendices}

\bibliographystyle{utphys}
\bibliography{reference2}

\end{document}